\documentclass[twocolumn,superscriptaddress,floatfix,preprintnumbers,prd ,nofootinbib,hyperref]{revtex4-2}
\usepackage[colorlinks=true,breaklinks=true]{hyperref}
\usepackage[utf8]{inputenc}
\hypersetup{allcolors=[rgb]{0.0 0.0 0.6},linkcolor=[rgb]{0.75 0.05 0.05}}
\usepackage{amsmath,amssymb}
\usepackage{mathtools}
\usepackage[normalem]{ulem}
\usepackage[dvipsnames]{xcolor}
\usepackage{float}
\usepackage{graphicx}
\usepackage{epsfig}
\usepackage{tensor}
\usepackage{letltxmacro}
\LetLtxMacro{\oldcite}{\cite}
\renewcommand{\cite}[1]{\mbox{\oldcite{#1}}}
\usepackage[caption = false]{subfig}

\usepackage{orcidlink}

\long\def\exclude#1{}

\DeclareMathOperator{\cm}{cm}

\DeclareMathOperator{\eV}{eV}

\DeclareMathOperator{\keV}{keV}

\DeclareMathOperator{\ms}{ms}

\DeclareMathOperator{\km}{km}
\DeclareMathOperator{\m}{m}

\DeclareSymbolFont{starfontsym}{OT1}{sts}{m}{n}
\DeclareMathSymbol{\mathTerra}{\mathord}{starfontsym}{76}

\newcommand{\beq}{\begin{equation}}
\newcommand{\eeq}{\end{equation}}

\def\ga{\,\,\raise0.14em\hbox{$>$}\kern-0.76em\lower0.28em\hbox
{$\sim$}\,\,}

\long\def\exclude#1{}

\allowdisplaybreaks

\bibpunct{[}{]}{,}{n}{}{,}

\hypersetup{
    colorlinks=true,
    citecolor=[rgb]{.1, .7, .6},
    linkcolor=[rgb]{.2, .55, .95},
    filecolor=magenta,
    urlcolor=[rgb]{.1, .7, .6},
}

\newcommand{\LL}{\mathcal{L}}

\newcommand{\OO}[1]{\mathcal O \left( #1 \right)}
\newcommand{\UU}{\mathcal U}
\newcommand{\EE}{\mathcal{E}}
\newcommand{\PP}{\mathcal{P}}

\newcommand{\spatial}[1]{\prescript{(3)}{}{#1}}
\newcommand{\tder}{\partial_{t}}
\newcommand{\abs}[1]{\lvert #1 \rvert}

\newcommand{\nel}{n_{\scriptscriptstyle{\rm EL}}}
\newcommand{\snc}{\sin{\chi_0} \,}

\newcommand{\dA}{A'}
\newcommand{\dE}{E'}
\newcommand{\dB}{B'}
\newcommand{\dF}{F'}
\newcommand{\da}{a'}
\newcommand{\dvarphi}{\varphi'}

\begin{document}

\preprint{CERN-TH-2024-179}

\title{Beyond The Standard Model electrodynamics in the time domain}

\author{Fabrizio Corelli}
\affiliation{Dipartimento di Fisica, ``Sapienza'' Universit\`a di Roma \& Sezione INFN Roma1, Piazzale Aldo Moro
5, 00185, Roma, Italy}

\author{Enrico Cannizzaro}
\affiliation{CENTRA, Departamento de F\'{\i}sica, Instituto Superior T\'ecnico -- IST, Universidade de Lisboa -- UL, Avenida Rovisco Pais 1, 1049 Lisboa, Portugal}

\author{Andrea Caputo} 
\affiliation{Department of Theoretical Physics, CERN, Esplanade des Particules 1, P.O. Box 1211, Geneva 23, Switzerland}

\author{Paolo Pani}
\affiliation{Dipartimento di Fisica, ``Sapienza'' Universit\`a di Roma \& Sezione INFN Roma1, Piazzale Aldo Moro
5, 00185, Roma, Italy}


\begin{abstract}
Many motivated extensions of the standard model include new light bosons, such as axions and dark photons, which can mix with the ordinary photon. This latter, when in a dilute plasma, can be dressed by an effective plasma mass. If this is equal to the mass of the new degree of freedom, then a resonance takes place and the probability of transition between states is enhanced. This phenomenon of resonance conversion is at the very basis of multiple probes of dark matter, and it is typically studied within the so-called Landau-Zener approximation for level crossings. This latter is known to break down in a variety of scenarios, such as multiple level crossings or when the de Broglie wave length of the new boson is comparable to the scale over which the background plasma varies.
We develop a flexible code adopting a 3+1 formalism in flat spacetime to perform non-linear simulations of systems with photons and new ultralight bosons in the presence of plasma. Our code currently allows to evolve one-dimensional systems, which are the ones of interest for this first study, but can be easily extended to treat three-dimensional spaces, and can be adapted to describe a plethora of realistic astrophysical and cosmological situations. Here we use it to study the breakdown of the Landau-Zener approximation in the case of multiple level crossings and when the slowly-varying plasma approximation ceases to be valid. In this first paper we detail our code and use it to study non-turbulent plasma, where small scale fluctuations can be neglected; their treatment will be considered in an upcoming publication. 
\end{abstract}

\maketitle

\section{Introduction}

The phenomenon of energy levels crossing in two-level systems has been studied extensively for decades. Almost 100 years ago, Landau and Zener~(LZ) examined the crossing of energy levels in the optical transitions of molecules~\cite{Landau:1932vnv} and between polar and homopolar molecular states~\cite{Zener:1932ws} under certain approximations. Thirty years later, Parke applied this concept to the survival probability of electron neutrinos in a resonant oscillation region~\cite{Parke:1986jy}, coining the term ``LZ transition probability". 

Today, the LZ approximation has been pivotal in developing many important probes of bosonic dark matter candidates, such as dark photons and axions. Both dark photons and axions can couple to the standard model photon, allowing for oscillations between different states. A propagating dark photon can oscillate into an ordinary photon and vice versa~\cite{Caputo:2020bdy,Caputo:2020rnx,An:2020jmf,An:2022hhb,An:2023wij, McDermott:2019lch, Mirizzi:2009iz, Beadle:2024jlr, Bolton:2022hpt}, and the same can happen for axions, albeit requiring an external magnetic field~\cite{Pshirkov:2007st, Battye:2019aco, battye2021robust, millar2021axionphotonUPDATED, Hook:2018iia, Foster:2020pgt, Gines:2024ekm, Tjemsland:2023vvc, Witte:2021arp, McDonald:2023shx, Mirizzi:2009nq, McDonald:2023ohd, Battye:2023oac, Battye:2021xvt}. Furthermore, in both cases the presence of a dilute plasma can enhance the conversion probability analogously to the neutrino case~\cite{Parke:1986jy}. Indeed, if the density of the plasma is such that the plasma frequency --~which acts as an effective mass for transverse electromagnetic excitations~-- matches the mass of the new boson, then a resonant conversion takes place and the probability of oscillation from one state to another is greatly boosted. Typically, this transition is treated within the approximation of the LZ approach, however the latter is known to break down in certain circumstances, such as in the presence of multiple level crossings or when plasma density variations occur on scales comparable to the de Broglie wavelength of the new boson. Consequently, most studies assume a slowly varying background and well-separated (or unique) level crossings. These assumptions, however, are likely to fail in many realistic astrophysical and cosmological contexts. Thus, it is of paramount importance to develop a fully general treatment of the resonant conversion phenomenon. 

A significant advancement in this direction was recently made with the introduction of a numerical scheme to study axion-photon mixing in strongly magnetized plasmas~\cite{Gines:2024ekm}, wherein the axion-Maxwell system of equations were numerically solved in a discretized spatial domain with a varying background. The results are in agreement with previous analytical derivations~\cite{McDonald:2023ohd} and were further validated in~\cite{McDonald:2024uuh}, demonstrating that the probability can be accurately reconstructed using numerical schemes.

In this work, we develop a flexible code adopting a 3+1 formalism in flat spacetime to perform real-time simulations for a system with photons and new light bosons, in the presence of an external plasma. As a first application we focus on the case of dark photons, although our code can be easily adapted to axion-photon oscillations as well. We verify the accuracy of the LZ approximation in scenarios with slowly varying plasmas and then quantify its failure in describing the conversion probability when multiple crossings are present or when the plasma varies rapidly. Our approach  presents a number of differences with that of Ref.~\cite{Gines:2024ekm}. While the latter works in the frequency domain, our approach provides a full time-domain evolution of the systems, allowing to study with ease also non-stationary systems. Moreover, while in this work we focus on flat spacetime, this scheme allows studying the systems in a generic spacetime. This could be relevant if one considers e.g. axion-photon mixing around neutron stars or conversion in cosmological scenarios.

Henceforth we use the metric signature $\{-, +, +, +\}$, and adopt rationalized Heaviside units with $\hbar=c=1$.

\section{Setup}  \label{sec:setup}
Our code can be applied to any new bosons which mixes with ordinary photons in a plasma. For simplicity, we focus here on the case of dark photons, but everything can be easily extended to the case of axions. Therefore, the Lagrangian density we consider can be expressed as~\cite{Holdom:1985ag, Fayet:1990wx}
\begin{align}
    \LL &= -\frac{1}{4} F_{\alpha\beta} F^{\alpha\beta} - \frac{1}{4} \dF_{\alpha\beta} \dF^{\alpha\beta} - \frac{\mu^2}{2} {\dA}_\alpha \dA^\alpha \notag \\
        &+ \frac{\snc}{2} \dF_{\alpha\beta} F^{\alpha\beta} - J^\alpha A_\alpha,
    \label{eq:Lagrangian}
\end{align}
where $A_\alpha$ and $\dA_\alpha$ are the photon and the dark photon, respectively, while $F_{\alpha\beta}$ and $\dF_{\alpha\beta}$ their field strenghts; $\mu$ is the mass of the dark photon, and $J^\alpha$ is the 4-current associated with the plasma fluid; lastly, $\chi_0$ is the mixing angle. 

In order to make the equations more tractable, it is convient to perform a field redefinition that removes the mixing  term present in \eqref{eq:Lagrangian}. The redefinition of the visible EM field $A_\alpha \to A_\alpha + \sin \chi_0 \dA_\alpha + \OO{\sin^2 \chi_0}$ leads to the Lagrangian in the so-called mass-basis~\cite{Fedderke:2021aqo}:
\begin{align}
    \LL_{\rm mass} &=  -\frac{1}{4} F_{\alpha\beta} F^{\alpha\beta} - \frac{1}{4} \dF_{\alpha\beta} \dF^{\alpha\beta} - \frac{\mu^2}{2} {\dA}_\alpha \dA^\alpha \notag \\
                   &- J^\alpha (A_\alpha + \snc \dA_\alpha),
    \label{eq:MassBasisLagrangian}
\end{align}
where we neglected terms of order $\sin^2 \chi_0$. From the Lagrangian density \eqref{eq:MassBasisLagrangian} we can derive the following field equations for the photon and the dark photon:
\begin{align}
    \nabla_\alpha F^{\alpha \beta} &= J^\beta \label{eq:FieldEM}, \\
    \nabla_\alpha \dF^{\alpha \beta} &= \snc \, J^\beta + \mu^2 \dA^{\beta} \label{eq:FieldDarkEM}.
\end{align}
For the plasma instead we will consider a cold electron fluid, and therefore we will neglect the pressure term. The force term will receive contributions from both the ordinary photon and the dark photon, resulting in the equations
\begin{align}
    u^\beta \nabla_\beta u^\alpha &= \frac{e}{m_e} \left( F^{\alpha\beta} + \snc \dF^{\alpha\beta} \right) u_\beta , \label{eq:PlasmaMomentum}\\
    \nabla_\alpha (n_e u^\alpha) &= 0  \label{eq:ContinuityEquation},
\end{align}
where $u^\mu$ is the 4-velocity of the fluid, while $m_e$ and $e$ are the mass and charge of the electron.

Another relevant basis can be obtained from the original one by performing the transformation $\dA^\alpha \to \dA^\alpha + \snc A^\alpha + \OO{\sin^2 \chi_0}$, and it is called interaction basis. After performing the field redefinition in \eqref{eq:Lagrangian} the Lagrangian density reads~\cite{Fedderke:2021aqo}:
\begin{align}
    \LL_{\rm interaction} &=  -\frac{1}{4} F_{\alpha\beta} F^{\alpha\beta} - \frac{1}{4} \dF_{\alpha\beta} \dF^{\alpha\beta} - \frac{\mu^2}{2} {A'}_\alpha \dA^\alpha \notag \\
                          &- \mu^2 \snc \dA_\alpha A^\alpha - J^\alpha A_\alpha
    \label{eq:InteractionBasisLagrangian}
\end{align}
In this basis, the hidden field is sterile, and does not interact with charged particle. Thus, the plasma momentum equation is the standard Lorentz force.
The interaction basis will be used when comparing our numerical results with the analytical predictions, as it is the basis in which the LZ formula is derived.
\subsection{Analytical derivation of the conversion probability}
Before delving in our numerical framework, we briefly elucidate the standard analytical procedure leading to the LZ formula for resonant conversion, following closely Refs.~\cite{PhysRevD.37.1237, An:2020jmf, An:2023mvf}.
\subsubsection{Framework}
We study  the propagation of plane waves in the interaction basis in the presence of an inhomogeneous plasma. For this work, we limit ourselves to situations which can be described by a 1+1 set-up; the field equations for the transverse fields can thus be expressed in the following form: 
\begin{equation}
    \Big[-\frac{\partial^2}{\partial_t^2}+\frac{\partial^2}{\partial_z^2}- 
    \begin{pmatrix}
    \omega_{\rm p}^2 & \snc\mu^2\\
     \snc \mu^2 & \mu^2 \\
    \end{pmatrix}\Big]
    \begin{pmatrix}
    A_T \\
    A'_T\\
    \end{pmatrix}=0 \, ,
\end{equation}
where $\omega_p = \sqrt{n_{e} e^2 / m_e}$ is the plasma frequency.
We are interested in monochromatic waves, i.e., solutions with fixed frequency $\omega$. Defining $k_z=\sqrt{\omega^2-\mu^2}$, we can search for solutions in the form $A_T(z,t)=\Tilde{A}_T(z)e^{i(\omega t - k_z z)}$ with an analogous ansatz for the dark photon field. 
Analytical solutions for this system can be obtained by assuming that the plasma frequency varies slowly with respect to $k$. This implies $|\partial_z \Tilde{A}_T(z)|\ll k_z |\Tilde{A}_T(z)|$ and same for $\Tilde{A}'_T$.
Under this assumption, the equation of motion of the transverse components of the fields can be recast into a  Schrödinger-like equation by adopting a WKB approximation. The latter reads $i\partial_z \bold{A}=(H_0+H_1)\bold{A}$, where $\bold{A}=(\Tilde{A}_T, \Tilde{A}'_T)$ encapsulates the  fields amplitudes while the Hamiltonian was split in a diagonal and off-diagonal part:
\begin{equation}
H_0= \begin{pmatrix}
\Delta-\Delta_{A'} & 0\\
0 & 0
\end{pmatrix} \ 
H_1= \snc \begin{pmatrix}
0 &  \Delta_{A'}\\
\Delta_{A'} & 0 
\end{pmatrix}  \, ,
\end{equation}
where:
\begin{equation}
    \Delta=-\omega_{\rm p}(z)^2/2 k_z  \ \ {\rm and} \ \ \Delta_{A'}=-\mu^2/ 2 k_z \, .
\end{equation}
We can solve this equation perturbatively since $\snc\ll 1$, relying on standard methods of quantum mechanics perturbation theory.
We define the free-evolution operator,
\begin{equation}
    \mathcal{U}(z)=\text{exp}\Big[-i\int_{z_i}^z H_0(z')dz'  \Big]\,,
\end{equation}
and switch to the interaction picture where we define $\bold{A}_{\text{int}}=\mathcal{U}^\dag \bold{A}$ and $H_{\text{int}}=\mathcal{U}^\dag H_1 \mathcal{U}$, so that the Schrödinger equation becomes $i\partial_z \bold{A}_{\text{int}}=H_{\text{int}}\bold{A}_{\text{int}}$.
Explicitly, $H_{\text{int}}$ can be expressed as
\begin{equation}
    H_{\text{int}}= \snc \Delta_{A'}
\begin{pmatrix}
0 & e^{i \phi(z)}\\
e^{-i \phi(z)} & 0 
\end{pmatrix}\, ,
\end{equation}
where we defined $\phi(z)=\int_{z_i}^z \Big(\Delta(z')-\Delta_{A'}\Big) dz'$.
We can thus define the evolution of the state in the interaction picture $\bold{A}_{\text{int}}(z)=\text{exp}\Big[-i\int_{z_i}^z H_{\rm{int}}(z')dz'  \Big]\bold{A}_{\text{int}}(z_i)$ and switch back to the Schrödinger picture using $\bold{A}=\mathcal{U}\bold{A}_{\text{int}}$. To first order in $\snc$ this yields, up to an overall phase,
\begin{equation}
    \bold{A}(z)=
    \begin{pmatrix}
1 & -i c_+\\
-i e^{i \phi}c_-  & e^{i \phi} 
\end{pmatrix}\bold{A}(z_i)\, ,
\end{equation}
where we defined $c_\pm=\snc \Delta_{A'}\int_{z_i}^z e^{\pm i \phi(z')dz'}$. The conversion probability can now be simply computed as the scalar product squared between the two states respectively at $z,z_i$. As such, it reads:
\begin{equation}\label{eq:probintegral}
    P_{A \leftrightarrow A'}=\Big|\int_{z_i}^z dz' e^{i \phi(z')}\snc \Delta_{A'}\Big|^2 \,.
\end{equation}
\subsubsection{Stationary phase approximation}
In general, the integral in Eq.~\eqref{eq:probintegral} oscillates rapidly, yet it varies more slowly near stationary points with $\phi'(z)=0$. One can therefore assume that far away from stationary points oscillations interfere destructively and average to zero, so that the main contribution to the integral is near resonant points $z_n$, where $\omega_p(z_n)=\mu$ so that $\phi'(z_n)=0$. We can therefore evaluate the integral analytically using the stationary phase approximation around resonances. Expanding the phase to second order, we obtain:
\begin{equation}\label{eq:LZformula}
    P_{A \leftrightarrow A'}=\sum_{z_n} \snc^2 \Delta_{A'}^2 \frac{2 \pi}{\abs{\phi''(z_n)}}\,,
\end{equation}
where the sum is over all resonant points and we ignored interference effects between different resonances. This is the LZ formula for resonant conversion. Summing up, the latter was obtained under two major assumptions: i) an underlying assumption of the whole computation was the WKB approximation, implying that the plasma parameters vary slowly in space; and ii) the stationary phase approximation requires that saddle points are not degenerate. In the rest of this paper, we will supersede both these assumptions.

\subsubsection{The case of coalescing saddle points}
The standard stationary phase approximation breaks down whenever saddle points coalesce in the vicinity of a local extremum of the plasma frequency~\cite{Brahma:2023zcw}. This is because around this critical point $z_c$, we have $\phi''(z_c)\propto\omega_{\rm p}'(z_c)=0$, so that the leading contribution due to the saddle points diverges.
In this case, one can rely on the cubic approximation. We can expand the phase around the extremal point to third order, assuming two saddle points are close to it. This yields 
\begin{equation}
    \phi(z)\approx \phi(z_c)+\phi'(z_c)(z-z_c)+\frac{\phi'''(z_c)}{3!}(z-z_c)^3\,,
\end{equation}
so that the integral~\eqref{eq:probintegral} can be solved analytically. The result reads
\begin{equation}
    P_{A \leftrightarrow A'}=4 \pi^2\left|\Big(\frac{2}{\phi'''(z_c)}\Big)^{\frac{1}{3}}\snc \Delta_{A'}A_i(\zeta)\right|^2
    \label{eq:CubicApproximation}
\end{equation}
where we defined $\zeta=\Big(2/\phi'''(z_c)\Big)^{1/3} \phi'(z_c)$ and introduced the Airy function,
\begin{equation}
    A_i(x)=\frac{1}{2\pi}\int e^{i(x v+v^3/3)}dv.
\end{equation}
Alternatively, Ref.~\cite{Brahma:2023zcw} also considered a so-called transitional Airy approximation. The latter is obtained by evaluating the integral~\eqref{eq:probintegral} using a uniform approximation and taking the coalescing limit. This approximation gives a further correction with respect to a standard cubic approximation (see~\cite{CONNOR, beuc, beuc2}):
\begin{align} \label{eq:TransitionalAiryApproximation}
    P_{A \leftrightarrow A'}=&4 \pi^2\Big|\Big(\frac{2}{\phi'''(z_c)}\Big)^{\frac{1}{3}}\snc \Delta_{A'}\Bigg(A_i(\zeta) \\ &\nonumber -i \Big(\frac{2}{\phi'''(zc)}\Big)^{\frac{1}{3}}\Big(\frac{\phi''''(z_c)}{6 \phi'''(z_c)} \Big) {A'}_{i}(\zeta))\Bigg)\Big|^2
\end{align}
In the following, we will scrutinize the validity of these approximations by comparing them with our fully-numerical results.

\subsection{3+1 decomposition of the field equations}  \label{sec:fieldequations}

Following Ref.~\cite{Cannizzaro:2023ltu}, we will write the evolution equations performing a 3+1 decomposition. From the practical point of view, this consists in foliating the spacetime using a set of spacelike hypersurfaces $\Sigma_t$, over which the coordinate time assumes a constant value. Then the field equations are projected onto the hypersurfaces and onto the vector $n^\mu$ orthogonal to them, and a set of evolution equations and constraints is obtained. In this context, we will use Greek letters to refer to spacetime indices, and Latin letters for the components of the tensors on the 3-dimensional hypersurfaces. Furthermore, we will use the vector notation to identify 3-vectors: $\vec V = (V^1, V^2, V^3)$. In our specific case, since we consider a flat background, $ds^2 = \eta_{\alpha\beta} dx^\alpha dx^{\beta}$, the normal vector is given by $n^\alpha = (1, 0, 0, 0) = (1, \vec 0)$; additionally, due to our choice of the signature, the contravariant and covariant components of a 3-vectors are the same, $V^i = V_i$.

Let us start with some necessary definitions. Introducing the dual of the electromagnetic tensor $F^{*\alpha\beta} = -\frac{1}{2} \epsilon^{\alpha\beta\lambda\sigma}F_{\lambda\sigma}$, we can define the electric and magnetic field as \cite{Alcubierre:2009ij}
\begin{equation}
    E^\alpha = -n_\beta F^{\beta\alpha}, \qquad B^{\alpha} = -n_\beta F^{*\beta\alpha}.  \label{eq:EBdef}
\end{equation}
Note that both $E^\alpha$ and $B^\alpha$ are on the 3-surface $\Sigma_t$, and therefore we have $E^\alpha = (0, \vec E)$, $B = (0, \vec{B})$. With the definitions \eqref{eq:EBdef}, and introducing the Levi-Civita tensor on the 3-surfaces $\spatial{\epsilon}^{\alpha\beta\sigma} = n_\lambda \epsilon^{\lambda\alpha\beta\sigma}$, we can rewrite $F^{\alpha\beta}$ in terms of the electric and magnetic fields as
\begin{equation}
    F^{\alpha\beta} = n^\alpha E^\beta - n^\beta E^\alpha + \spatial{\epsilon}^{\alpha\beta\sigma} B_\sigma. \label{eq:Fdecomp}
\end{equation}
Then, we can decompose the 4-potential of the electromagnetic field as
\begin{equation}
    A^\alpha = \varphi n^\alpha + a^\alpha, \label{eq:Adecomp}
\end{equation}
where $\varphi = -n_\alpha A^\alpha$ and $a^\alpha = \tensor{h}{^\alpha_\beta} A^\beta$ are the scalar and vector potentials, respectively, and $\tensor{h}{^\alpha_\beta}$ is the operator that performs the projection onto the spatial 3-surfaces. An analogous decomposition can be obtained for the dark electromagnetic field by performing the same steps.

To decompose the 4-velocity of the electrons in the plasma, we introduce its projections $\Gamma = -n_\alpha u^\alpha$, and $\spatial{u}^\alpha = \tensor{h}{^\alpha_\beta} u^\beta$. If we additionally define $\UU^\alpha$ in such a way that $\spatial{u}^\alpha = \Gamma \UU^\alpha$, we can arrive at
\begin{equation}
    u^\alpha = \Gamma \bigl(n^\alpha + \UU^\alpha \bigr). \label{eq:udecomp}
\end{equation}
Note that $\UU^\alpha = (0, \vec \UU)$ since it lies on the spatial 3-surfaces.

Lastly, we need to define and decompose the electromagnetic 4-current. In general, it can be decomposed as 
\begin{equation}
    J^\alpha = -\rho n^\alpha + \spatial{J}^\alpha,
\end{equation}
where $\rho = n_\alpha J^\alpha$ is the charge density, and $\spatial{J}^\alpha = \tensor{h}{^\alpha_\beta} J^\beta$ is the 3-current. In our specific case the current is given by the plasma fluid, so that it receives contributions from electrons and ions, $J^\alpha = J_{\rm (ions)}^\alpha + J_{\rm (e)}^\alpha$. In this work we will assume ions to be at rest, thanks to the fact that their mass is considerably larger than the mass of electrons. As a result, we can write $J_{\rm (ions)}^\alpha = -\rho_{\rm (ions)} n^\alpha$, where $\rho_{\rm (ions)}$ is the charge density of ions. For electrons instead we have $J_{\rm (e)}^\alpha = - e n_e u^\alpha = -e n_e \Gamma \bigl(n^\alpha + \UU^\alpha \bigr) = -e \nel \bigl(n^\alpha + \UU^\alpha\bigr)$, where we defined $\nel = n_e \Gamma$. As a result we have that the charge density and current of the plasma are respectively given by
\begin{equation}
    \rho = \rho_{\rm (ions)} + e \nel, \qquad \spatial{J}^\alpha = - e \nel \UU^\alpha.
    \label{eq:EMChargeCurrent}
\end{equation}

With these definitions at hand we can write the system of evolution equations and constraints. For simplicity, we will perform this operation directly in the flat background case, although the extension to a fixed curved spacetime is straightforward. 

To obtain the equations for the electromagnetic fields, we follow the procedures described in Ref.~\cite{Alcubierre:2009ij}. By projecting Eq.~\eqref{eq:FieldEM} onto the normal vector $n^\alpha$, we obtain the Gauss law,
\begin{equation}
    \partial_i E^i = \rho,
    \label{eq:Gauss}
\end{equation}
while projecting on the spatial 3-surface we obtain an evolution equation for $\vec E$,
\begin{equation}
    \tder E^i = \bigl(\vec \partial \times \vec B \bigr)^i + \spatial{J}^i.
    \label{eq:EiEvol}
\end{equation}
The evolution equation for the 3-potential $a^i$ can be obtained from the definition of the electric field in Eq.~\eqref{eq:EBdef}, and it reads
\begin{equation}
    \tder a_i = - E_i - \partial_i \varphi,
    \label{eq:AiEvol}
\end{equation}
while the evolution equation for the scalar potential can be derived from the gauge choice. Using the Lorenz gauge we obtain
\begin{equation}
    \tder \varphi = -\partial_i a^i.
    \label{eq:varphiEvol}
\end{equation}
Finally, we should provide an explicit expression for the magnetic field $\vec B$:
\begin{equation}
    B^i = \bigl( \vec \partial \times \vec a \bigr)^i.
    \label{eq:Bcurla}
\end{equation}

Analogous operations can be performed starting from Eq.~\eqref{eq:FieldDarkEM}, to obtain the evolution equations and constraints for the dark photon field. From the practical point of view it is sufficient to replace $J^\alpha \to \snc J^\alpha + \mu^2 \dA^\alpha$ in Eqs.~\eqref{eq:Gauss}-\eqref{eq:Bcurla} obtaining
\begin{align}
    \partial_i \dE^i &= \snc \rho - \mu^2 \dvarphi,                                                      \label{eq:DarkGauss} \\
    \tder \dE^i &= \bigl(\vec \partial \times \vec \dB \bigr)^i + \snc \spatial{J}^i + \mu^2 \da^i,      \label{eq:DarkEiEvol} \\
    \tder \da_i &= - \dE_i - \partial_i \dvarphi,                                                        \label{eq:DarkAiEvol}  \\
    \tder \dvarphi &= -\partial_i \da^i,                                                                 \label{eq:DarkvarphiEvol}  \\
    \dB^i &= \bigl( \vec \partial \times \vec \da \bigr)^i.                                              \label{eq:DarkBcurla}
\end{align}
However, unlike for the ordinary photon, here the Lorenz condition $\nabla_\alpha \dA^\alpha$ is not a result of a gauge choice, but an identity obtained by taking the 4-divergence of Eq.~\eqref{eq:FieldDarkEM}.

The procedures to obtain the evolution equations for the plasma fluid are described in Appendix A of Ref.~\cite{Cannizzaro:2023ltu}, in which a similar model without the dark photon was studied. However, given the structure of Eq.~\eqref{eq:PlasmaMomentum}, we can obtain the evolution equations from Ref.~\cite{Cannizzaro:2023ltu}, simply performing the substitutions
\begin{equation}
    \vec E \to \vec E + \snc \vec \dE, \qquad \vec B \to \vec B + \snc \vec \dB.
\end{equation}
The result is
\begin{align}
    \tder \Gamma &= - \UU^i \partial_i \Gamma + \frac{e}{m_e} \bigl(E^i + \snc \dE^i \bigr) \UU_i,	\label{eq:GammaEvol} \\
	\tder \UU^i &= - \UU^j \partial_j \UU^i + \frac{1}{\Gamma} \frac{e}{m_e} \Bigl[ -\UU^i \bigl(E^j + \snc \dE^j \bigr) \UU_j \notag \\
                &+ E^i + \snc \dE^i + \bigl(\vec{\UU} \times \vec{B} \bigr)^i \notag \\
                &+ \snc \bigl(\vec{\UU} \times \vec{\dB} \bigr)^i \Bigr],	\label{eq:UUEvol} \\
    \tder \nel &= - \UU^i \partial_i \nel - \nel \partial_i \UU^i. 	\label{eq:NELEvol}
\end{align}
Finally, we have the constraint
\begin{equation}
	\Gamma^2(1 - \UU^i \UU_i) = 1\,,
	\label{eq:PlasmaConstraint}
\end{equation}
which comes from the normalization of the 4-velocity of the fluid, $u_\alpha u^\alpha = -1$.

\subsection{Numerical evolution}  \label{sec:numericalevolution}

We evolve $\vec{E}$, $\vec{a}$, $\varphi$ using Eqs.~\eqref{eq:EiEvol},~\eqref{eq:AiEvol},~\eqref{eq:varphiEvol} for the ordinary photon, and $\vec \dE$, $\vec \da$, $\dvarphi$ using Eqs.~\eqref{eq:DarkEiEvol},~\eqref{eq:DarkAiEvol},~\eqref{eq:DarkvarphiEvol} for the dark photon. For the plasma instead we evolve $\Gamma$, $\vec \UU$ and $\nel$ using Eqs.~\eqref{eq:GammaEvol},~\eqref{eq:UUEvol},~\eqref{eq:NELEvol}, respectively. The fields $\vec B$ and  $\vec \dB$ appearing in the equations are computed using Eqs.~\eqref{eq:Bcurla} and~\eqref{eq:DarkBcurla}, while their first derivatives are computed directly as second derivatives of the vector potentials $\vec a$ and $\vec \da$. 

The above evolution and constraint equations are general. For simplicity, in this work we consider systems that are homogeneous along the $xy$ plane, so that all the fields depend only on the $z$ coordinate and on time. Such assumption is consistent with the scenarios we are interested in studying, and allows us to perform the integration on a one-dimensional domain, which has the benefit of substantially reducing the computational cost compared to the full 3+1 case.

The numerical evolution is performed using the method of lines; spatial derivatives are computed with finite differences operators that satisfy the summation-by-parts~\cite{DelReySBPReview} property. This allows us not to impose boundary conditions, as summation-by-parts operators already compute the derivatives at the boundaries using only points in the interior of the numerical grid, without requiring the introduction of ghost zones. As for the accuracy, we use operators that guarantee fourth-order accuracy in the interior of the grid and second-order accuracy at the boundaries. The coefficients we used can be found in Appendix~C of Ref.~\cite{MATTSSON2004503}. For time integration we use the sixth-order accurate Runge-Kutta method, as we observed that the fourth-order accurate version introduces energy losses that are not negligible compared to the ones induced by the conversion of dark photons to photons. 

We evaluate the convergence of the code by checking the scaling with the resolution of the violation of the constraints in Eqs.~\eqref{eq:Gauss}, \eqref{eq:DarkGauss} and \eqref{eq:PlasmaConstraint}. The results of the tests we performed of our code are discussed in Appendix~\ref{app:convergence}.

\subsection{Initialization Procedure}  \label{sec:initialization}

\begin{figure}
    \centering
    \includegraphics[width=\columnwidth]{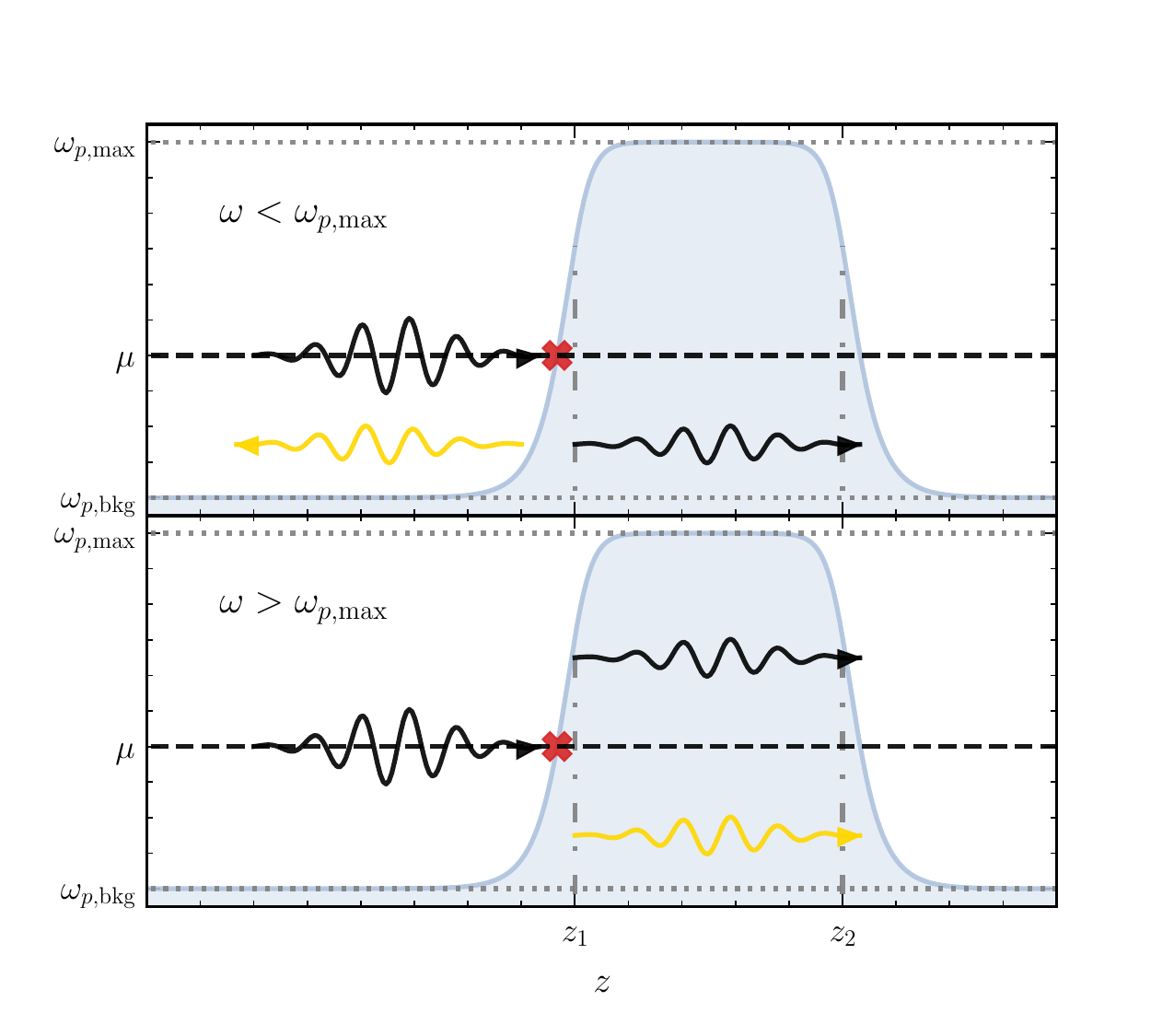}
    \caption{Schematic representation of the process we simulate. The blue-shaded area represents the plasma barrier, and the horizontal black dashed line is the dark photon mass. The resonant point is located where $\mu = \omega_p$, and is represented by a red cross. The dark photon and the photon wave packets are indicated with black and gold wavy arrows, respectively. In this work we consider a wave packet of the dark photon scattering off a plasma barrier. When it reaches the resonant point it undergoes the conversion process; the residual dark photon continues to propagate through the barrier, while the produced photon can be either reflected or transmitted depending on whether its frequency is lower of greater than the plasma frequency at the top of the barrier. These two cases are depicted in the upper and lower panel, respectively.}
    \label{fig:cartoon}
\end{figure}

The physical configuration we simulate is described by a wave packet of the dark photon scattering off a barrier of plasma. 
A schematic representation of the process is depicted in Fig.~\ref{fig:cartoon}, in which the blue-shaded area represents the plasma barrier, the red cross is the resonant point, where the plasma frequency is equal to the dark photon mass, while the gold and black wavy arrows represent the photon and the dark photon, respectively. When the dark photon reaches the resonant point it undergoes conversion; the photon produced can either be reflected by the plasma barrier if its frequency is smaller than the plasma frequency at the top of the barrier (upper panel), or transmitted through it in the opposite case (lower panel); the component of the dark photon that has not undergone conversion continues to propagate through the barrier. 

Our initial setup is similar to the one we considered in Ref.~\cite{Cannizzaro:2023ltu}. In particular, we use the same profile for the plasma density, which is 
\begin{align}
    n_0 = \nel(z, t = 0) &= n_{\rm bkg} + (n_{\rm max} - n_{\rm bkg}) \notag \\
                   &\times\Bigl[\sigma(z; W_1, z_1) + \sigma(z; -W_2, z_2) - 1\Bigr],
    \label{eq:InitialPlasmaDensity}
\end{align}
where $n_{\rm bkg}$ and $n_{\rm max}$ are the values of the plasma density on the background and on top of the barrier, respectively. The shape of the boundaries of the barrier is delineated by the sigmoid function $\sigma(z; W, z_0) = (1+e^{-W(z - z_0)})^{-1}$, with $z_{1,2}$ determining their position, and $W_{1,2}$ their steepness. Assuming the plasma to be initially neutral, we can then set the (constant) charge density of ions as $\rho_{\rm (ions)} = - e n_0$. As for the initial 4-velocity of the fluid, we set $\Gamma(z, t = 0) = 1$ and $\vec \UU(z, t = 0) = \vec 0$. The plasma profile we use schematically corresponds to the one shown in Fig.~\ref{fig:cartoon}, where we decided to directly represent the plasma frequency $\omega_p$.

The dark photon field is initialized as a wave packet with profile
\begin{align}
	\vec{\dE} &= A_E
	\begin{pmatrix}
		\cos[k_z(z - z_0)]\\
		\sin[k_z(z - z_0)]\\
		0
	\end{pmatrix}
	e^{-\frac{(z - z_0)^2}{2 \sigma^2}}, \\
	\vec{\dB} &= A_E \frac{k_z}{\omega}
	\begin{pmatrix}
		- \sin[k_z(z - z_0)]\\
		\cos[k_z(z - z_0)]\\
		0
	\end{pmatrix}
	e^{-\frac{(z - z_0)^2}{2 \sigma^2}},
	\label{eq:EMInitial}
\end{align}
where $k_z = \sqrt{\omega^2 - \mu^2}$. The electromagnetic field is instead initialized setting $\vec{E} = \vec{B} = 0$. It is worth underlying here that the initialization is performed in the mass basis, and the initial configuration actually contains a component of the ordinary photon with amplitude of order $A_E \snc$. On the other hand, the profiles in Eq.~\eqref{eq:EMInitial} coincide with the $\vec \dE$ and $\vec \dB$ components of the dark photon field of the original model \eqref{eq:Lagrangian}.

Once the profiles for $\vec{E}$, $\vec{B}$, $\vec{\dE}$, $\vec{\dB}$ have been set, we compute the potentials. 
Thanks to our assumed homogeneity on the $xy$ plane, we can obtain the profile of $\vec a$ by integrating Eq.~\eqref{eq:Bcurla} along the $z$ axis. This results into
\begin{align}
    a^x(z, t = 0) &= a^x(z = z_{+\infty}, t = 0) + \int_{z_{+\infty}}^z d \tilde z \, B^y(\tilde z, t = 0), \notag \\
    a^y(z, t = 0) &= a^y(z = z_{+\infty}, t = 0) - \int_{z_{+\infty}}^z d \tilde z \, B^x(\tilde z, t = 0), \notag \\
    a^z(z, t = 0) &= a^y(z = z_{+\infty}, t = 0),
\end{align}
where, $z_{+\infty}$ is the value of the $z$-coordinate on the right boundary of the numerical grid, and $a^i(z = z_{+\infty}, t = 0)$ are integration constants that we set to zero. For $\varphi$ instead we take advantage of the residual gauge freedom to impose $0 = \tder a_z(z, t = 0) = -E_z(z, t = 0) - \partial_z \varphi(z, t = 0)$, which can be inverted to give
\begin{equation}
    \varphi(z, t = 0) = \varphi(z = z_{+\infty}, t = 0) - \int_{z_{+\infty}}^z d \tilde z \, E_z(\tilde z, t = 0).
\end{equation}
Also in this case we set the integration constant $\varphi(z = z_{+\infty}, t = 0)$ to zero. The same formulas are used to obtain the scalar and vector potentials for the dark photon from the fields $\vec{\dE}$ and $\vec{\dB}$.

We compute the integrals numerically with the Simpson's rule, which guarantees fourth-order  accuracy. Since this procedure makes use of 3 grid points on each integration step, it is necessary to provide a way to initialize the computation that preserves the accuracy to fourth order. To circumvent this problem, we decided to perform a complete integration step over each grid step, by inserting fictitious grid points in the middle of the grid steps, and evaluating the integrands on them with the analytic expressions that the fields are initialized to. Furthermore, we perform the integration from the right boundary $z_{+\infty}$ to the left boundary $z_{-\infty}$, in order to reduce the numerical error in the proximity of the plasma barrier. Indeed, the profiles of the potentials $\vec{\da}$ and $\dvarphi$ are solely determined by the wave packet of the dark photon. As a result numerical error will accumulate in the region where the wave packet is situated, and will be larger after it than before it in the direction where the integration proceeds. Our strategy allows therefore placing the region where numerical errors are larger on the left of the wave packet, rather than on the right, where the plasma barrier is situated and where the conversion takes place.

\subsection{Choice of the grid parameters}  \label{sec:simparams}

Let us now briefly discuss the choice of the grid parameters. Given the different length scales that appear in our setup, this choice is particularly delicate. Indeed, the spatial grid step $\Delta z$ has to be small enough to resolve the smallest length scales over which the variation of the fields occurs. In our setup, given that we perform simulations in a regime in which the plasma is essentially stationary, in order to set $\Delta z$ we only have to evaluate the wave lengths of the photon and the dark photon.

We start our simulations with a wave packet of the dark photon with given frequency $\omega$. Its wavelength is therefore given by
$\lambda_{\rm DP} = 2 \pi / k_z = 2 \pi / \sqrt{\omega^2 - \mu^2}$. After the conversion the outgoing photon will have frequency $\omega$, and wave length $\lambda_{\rm P} = 2 \pi / \sqrt{\omega^2 - \omega_p^2}$, where $\omega_p$ is the plasma frequency in the region where the photon is situated. This means that the wave length of the photon assumes its minimum value outside the barrier, where $\omega_p = \omega_{p, \rm bkg} = \sqrt{n_{\rm bkg} e^2 / m}$, and $\lambda_{\rm P}^{\rm min} \approx 2 \pi / \omega$, as we always set $\omega \gg \omega_{p, \rm bkg}$. Furthermore, since we take $\mu > \omega_{p, \rm bkg}$, we have $\lambda_{\rm DP} > \lambda_{\rm P}^{\rm min}$. As a result $\lambda_{\rm P}^{\rm min}$ is the smallest physical length scale in the system. Given these considerations, we typically set a grid step of the order $\Delta z \lesssim \lambda_{\rm P}^{\rm min} / 40$, in order to be sure to correctly resolve the photon produced in the process we simulate.

As for the time step, we set $\Delta t = {\rm CFL} \times \frac{\Delta z}{v_p}$, where ${\rm CFL}$ is the Courant–Friedrichs–Lewy factor, and $v_p = \omega / k_{z}$ is the phase velocity of the dark photon. The value of the CFL factor that we use is $0.2$, as from tests we performed on our setups we observed that it does not lead to numerical instabilities, while producing an energy loss small enough for our purposes.

\section{Results}  \label{sec:results}

Before presenting our numerical results, it is worth noting that we will report distances and times in international units for the sake of convenience, although the numerical simulations are performed using natural units. Given that the equations are already derived in Heaviside units with $c = 1$, this is simply achieved by setting $e = \sqrt{4 \pi / 137}$, and $m_e = 511 \, \keV$. In all simulations we set the mixing angle to $\chi_0 = 10^{-4}$\; however, by performing a dedicated set of runs we have been able to check that the probability correctly satisfies the scaling with $\sin^2 \chi_0$, for $\sin \chi_0 \ll 1$.

\subsection{Testing the LZ formula in its regime of validity} \label{sec:results:LZ}

We start our analysis by performing a set of simulations in which the wave packet of the dark photon undergoes a single resonant conversion in the rising edge of the plasma barrier. For this purpose, we set the location of the left boundary of the plasma profile at $z_1 = 0 \, \km$, while the right boundary was placed at $z_2 = 10^5 \, \km$, far outside the numerical grid. The parameters $W_{1,2}$ that appear in Eq.~\eqref{eq:InitialPlasmaDensity} were set to $W_1 = W_2 = 0.01 \, \km^{-1}$, while the density parameters were set to $n_{\rm bkg} = 10^{-6} \, \cm^{-3}$ and $n_{\rm max} = 250 \, \cm^{-3}$, corresponding to plasma frequencies $\omega_{p, \rm bkg} = 3.70 \times 10^{-14} \, \eV$ and $\omega_{p, \rm max} = 5.86 \times 10^{-10} \, \eV$. In order to have a reference for a realistic physical environment, we have set the values of the spatial grid and electron densities to be close to their values within the ionosphere, as recently studied in Ref.~\cite{Beadle:2024jlr}. However, one can easily change these parameters to study different environments, such as the solar corona~\cite{An:2023mvf} or neutron star magnetospheres~\cite{Gines:2024ekm}.

The mass of the dark photon was set to $\mu = 10^{-10} \, \eV$, and we varied the frequency of the initial wave packet across the set, in order to consider different values for the group velocity. In particular, $\omega$ ranges from $1.005 \times 10^{-10} \, \eV$ to $3.2 \times 10^{-10} \, \eV$, resulting in group velocities that range approximately from $0.1$ to $0.95$. Note that we have $\omega < \omega_{p, \rm max}$ across the whole set, so that the photon produced after the conversion process can never penetrate the plasma barrier. The values of the parameter $\sigma$ appearing in Eq.\eqref{eq:EMInitial} ensure a good localization of the wave packet in terms of the wave number. In other words, we have $0.01 \lesssim \sigma_k/k_z \lesssim 0.1$, where  $\sigma_k = \sigma^{-1}$ is the width in terms of $k_z$. The initial localization of the packet, $z_0$, was chosen in such a way that its distance from the base of the plasma barrier was larger than $5 \sigma$, with a minimum of $500 \, \km$. The only exception is the case with $v_g = 0.1$, for which we set $z_0$ to ensure a distance $4 \sigma$ from the base of the barrier, in order to reduce the computational cost. The amplitude was set to $A_E = m_e \omega / (1000e)$  to ensure the linear regime of interaction with plasma, i.e. the regime in which the nonlinear terms in Eq.~\eqref{eq:UUEvol} are small, and the backreaction of the electromagnetic field on plasma is negligible. Indeed, nonlinear effects start becoming relevant when the amplitude of the electric field is $E_{\rm crit} \gtrsim m_e \omega /e$ \cite{1970PhFl...13..472K, PhysRevLett.64.2011}. In our setup we set only the amplitude of the dark photon wave packet, and the amplitude of the electric field produced after the conversion is related to it by the mixing parameter $\snc \approx 10^{-4}$. To be conservative and make sure that the system is always sufficiently far from the nonlinear regime, we decided to set $A_E$ to the critical threshold for the electric field, divided by a factor $1000$.

As for the numerical grid, the positions of the boundaries varied across the set, with larger numerical grids used for the simulations with smaller group velocity of the dark photon, due to the larger wave length and larger width of the wave packet. In general, we placed the boundaries in such a way that the conversion is completed before eventual signals reflected from the boundaries reach the resonant point. This results in having the left boundary of the grid at $-48000 \, \km \le z_{-\infty} \le - 7450 \, \km$, and the right boundary at $7150 \, \km \le z_{+\infty} \le 13000 \, \km$. The grid step and time step have been chosen using the criteria described in Sec.~\ref{sec:simparams}, while the total integration time was set in such a way that at the end of the simulation the wave packet of the dark photon was approximately at $z = -z_0$, i.e. at the same distance from the barrier as in the initial configuration, but inside it.

Unfortunately, we experienced a crash of the simulation with $v_g = 0.1$ due to instability appearing at the boundaries. By placing the $z_{-\infty} = -48000 \, \km$ we have been able to postpone the appearance of the instability (and the crash of the simulation) after the conversion process has completed.
It is worth mentioning, however, that this simulation is the most demanding one from the computational point of view. Indeed, a small group velocity translates in a large wave length, which requires a large width of the wave packet and the use of a large numerical grid as a result. Additionally, since $v_p = \omega / k_{\rm DP} = 1 / v_g = 10$, the time step $\Delta t$ has to be small compared to the case of larger $v_g$, in order to maintain the CFL factor to $0.2$. Lastly, the final time of the simulation has to be large, since the wave packet takes longer to propagate. All these factors simultaneously contribute in increasing the computational cost conspicuously.

\begin{figure*}[th]
    \centering
    \subfloat[Mass basis]{\label{fig:LZsnapshotsMass}\includegraphics[width=0.49\textwidth]{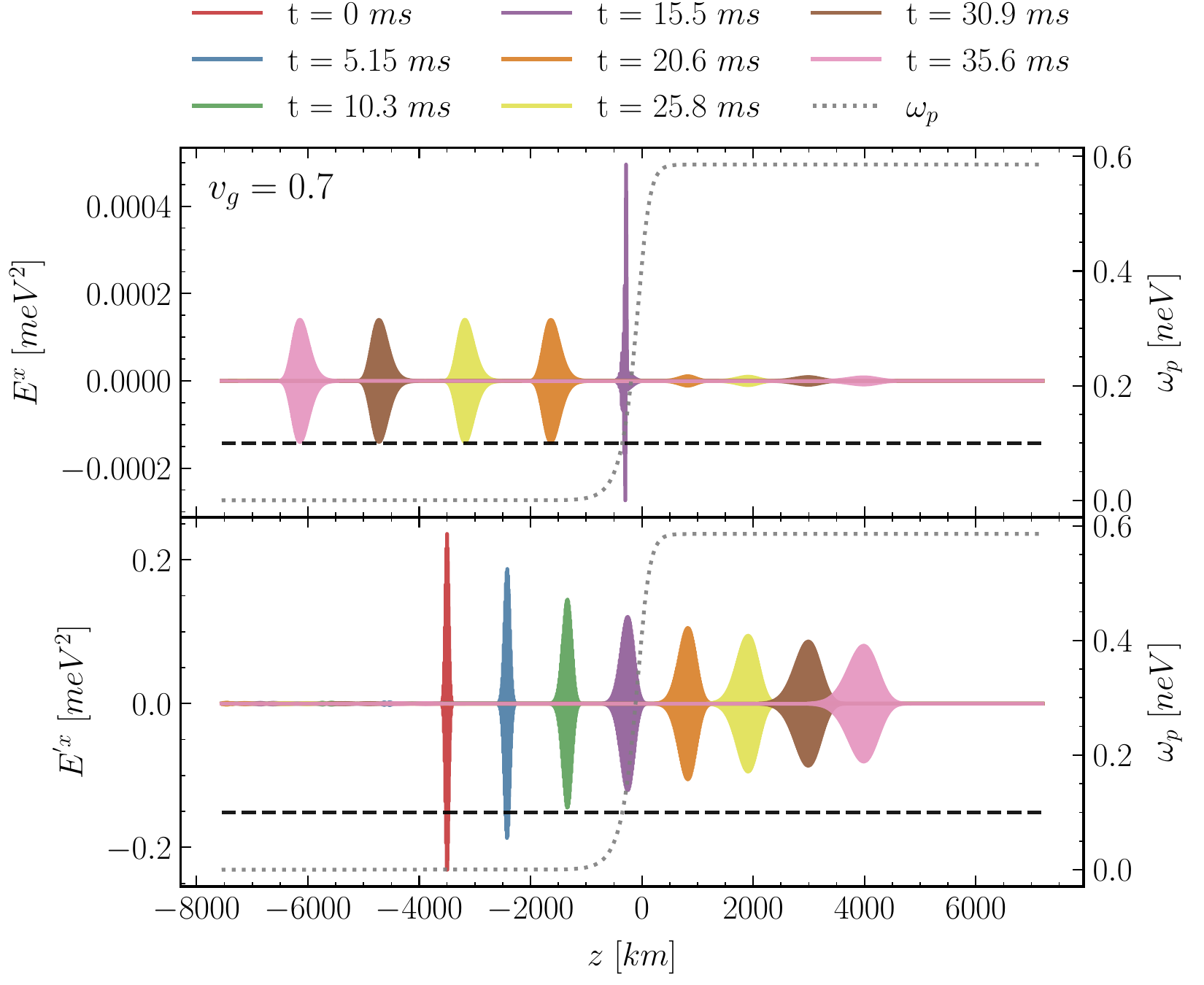}}
    \subfloat[Interaction basis]{\label{fig:LZsnapshotsInteraction}\includegraphics[width=0.49\textwidth]{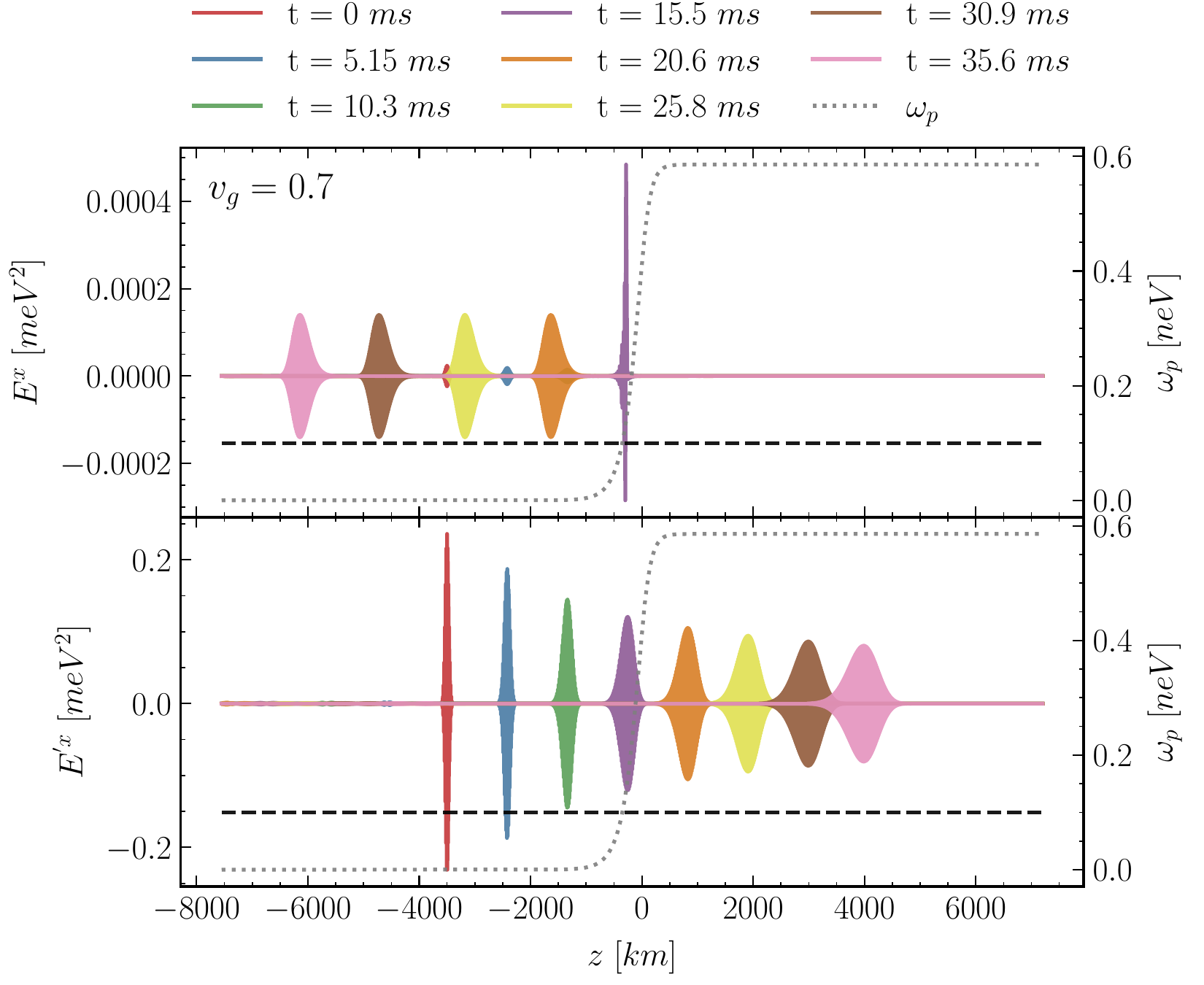}}\qquad
    \caption{Snapshots of the evolution for the simulation of the single conversion from dark photon to photon, in the regime of validity of the LZ formula, using both the mass basis (left) and the interaction basis (right). Here we show the case in which $v_g = 0.7$. In each of the two plots, solid lines denote the profile of $E^x$ and $\dE^x$ in the upper and lower panel, respectively, while the dotted line indicates the profile of the plasma density. The horizontal black dashed line indicates the value of the dark photon mass, $\mu = 10^{-10} \, \eV$, and the resonant point is located in its intersection with the dotted line. We can clearly observe the conversion taking place at the resonance, with the outgoing dark photon propagating inside plasma, and the produced photon being reflected towards the left, having a frequency smaller than $\omega_{p, \rm max}$. We highlight that the reduction of the wave-packet amplitude \textit{before} the resonant point is due to the natural dispersion that any finite-width wave-packet undergoes as it propagates.
    }
    \label{fig:LZsnapshots}
\end{figure*}

Some snapshots of a typical simulation are shown in Fig.~\ref{fig:LZsnapshotsMass} and Fig.~\ref{fig:LZsnapshotsInteraction}, in which we expressed the fields in the mass and interaction basis, respectively. The former is the eigenbasis in which the numerical integration is performed, while the latter has been obtain by performing the transformation according to the rule
\begin{align}
    {A_\alpha}^{\rm (int)} &= {A_\alpha}^{\rm (mass)} + \snc {\dA_\alpha}^{\rm (mass)} + \OO{\sin^2 \chi_0}, \notag \\
    {\dA_\alpha}^{\rm (int)} &= {\dA_\alpha}^{\rm (mass)} - \snc {A_\alpha}^{\rm (mass)} + \OO{\sin^2 \chi_0}.
    \label{eq:MassToInteractionBasis}
\end{align}

In both plots of Fig.~\ref{fig:LZsnapshots}, solid lines denote the profile of $E^x$ in the upper panel, and $\dE^x$ in the lower panel. The dotted line denotes the profile of the plasma density, and the horizontal black dashed line indicates the value of the dark photon mass. We can clearly see the conversion happening at the resonant point, which is where the dotted line intersects the horizontal dashed line. After the conversion the dark photon continues to propagate inside the plasma, while the photon cannot penetrate the barrier, having $\omega < \omega_{p, \rm max}$, hence it is reflected and propagates toward the left. Interestingly in Fig.~\ref{fig:LZsnapshotsMass}, a small wave packet of the photon appears to propagate inside the barrier. This is due to the fact that in the mass eigenbasis the electromagnetic field does not coincide with the standard model photon, and contains a component of the dark photon. The absence of photons after the resonant point is instead visible in the interaction basis (Fig.~\ref{fig:LZsnapshotsInteraction}). It is also worth noting that since in our initialization procedure, performed in the mass basis, we set a vanishing profile of the electromagnetic field, our simulations actually start with a small but nonvanishing contribution of the standard model photon, as it is visible from Fig.~\ref{fig:LZsnapshotsInteraction}.

To estimate the conversion probability we used the decrease in the energy of the dark photon across the resonance. In particular, if we call ${\mathcal E}_i$ and ${\mathcal E}_f$ the energy of the dark photon before and after the conversion, respectively, the following relation holds:
\begin{equation}
    {\EE}_f =  (1 - \PP){\EE}_i,
    \label{eq:DPEnergyLossLZ}
\end{equation}
where $\PP$ is the conversion probability. As a result we can obtain $\PP$ from our simulations as
\begin{equation}
    \PP = -\frac{\Delta \EE}{\EE_i} = \frac{\EE_i - \EE_f}{\EE_i}.
    \label{eq:LZSimulationProbability}
\end{equation}

In our 1+1 setup, we did not directly compute $\EE$, but rather the energy per unit surface which, apart from multiplicative factors, can be written as
\begin{equation}
    \overline \EE = \int_{z_{0}}^{z_{+\infty}} dz \, \bigl( \abs{\vec \dE}^2 + \abs{\vec \dB}^2 \bigr).
    \label{eq:DPEnergy}
\end{equation}
Then, the conversion probability can be computed using $\overline \EE$ in place of $\EE$, as both the surface term and the multiplicative factors that we dropped cancel out. The integral was performed numerically with the Simpson's rule, with $z_0$ and $z_{+\infty}$ being the initial position of the wave packet and the right boundary of the numerical domain of integration, respectively. This choice is motivated by the fact that the initial profile for the dark photon, Eq.~\eqref{eq:EMInitial} is not a fully consistent wave packet solution propagating towards the right. As a result a tiny component of the dark photon field propagates toward the left, and does not undergo the conversion process. Using $z_0$ as left boundary of the domain of integration allows us to exclude such component when evaluating the conversion probability $\PP$.

\begin{figure}
    \includegraphics[width = \columnwidth]{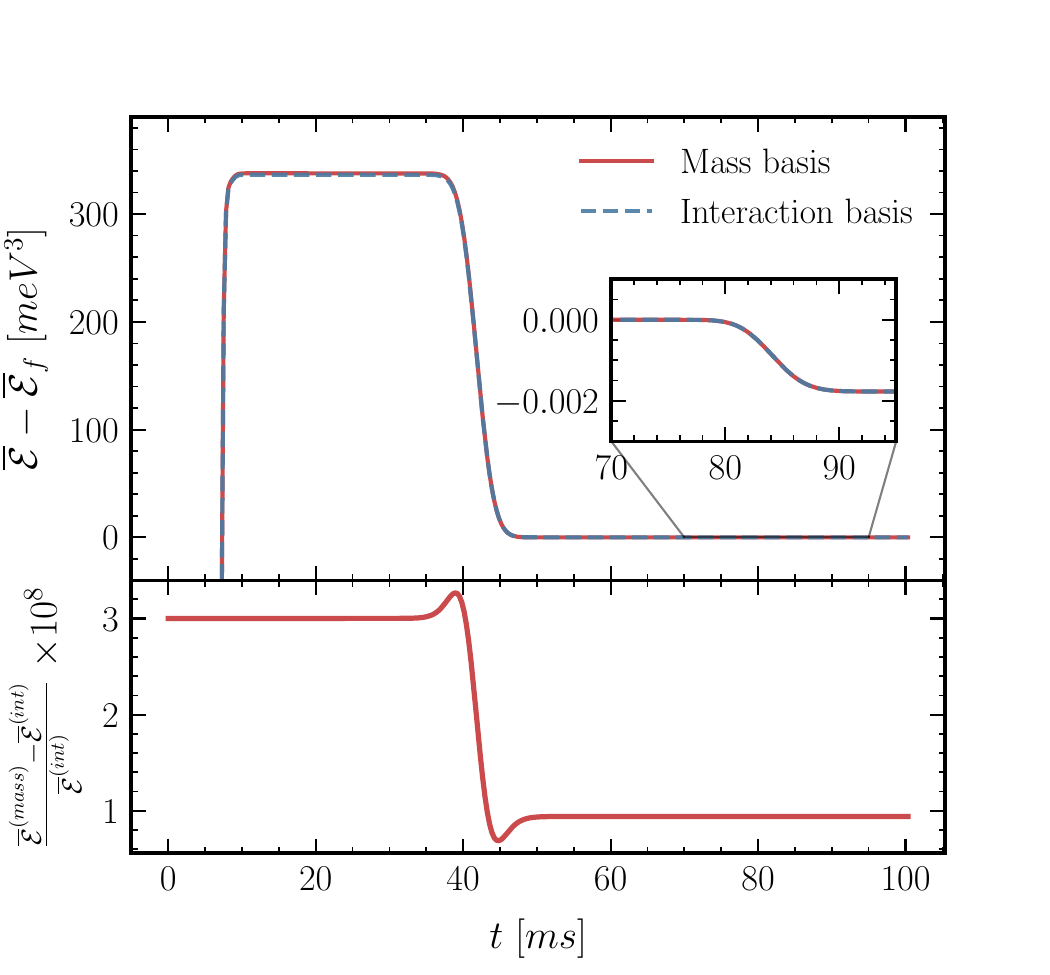}
    \caption{
    Upper panel: behavior of the energy per unit surface of the dark photon for the simulations with $v_g = 0.3$. We rescaled $\overline \EE$ by subtracting its value after the conversion. At early times the wave packet gradually enters in the integration domain, causing an increase in energy. At the time of the conversion $\overline \EE$ decreases in a clear step-like behavior, and it finally undergoes a small additional decrease due to fact that the component of the dark photon that is reflected by the barrier leaves the domain over which the energy is computed. This latter stage is shown in more detail in the inset. 
    Lower panel: relative discrepancy between the energy per unit surface computed in the mass basis and the interaction basis.
    \label{fig:DPEnergyLossLZ}
    }
\end{figure}

A typical behavior of the energy of the dark photon is shown in the top panel of Fig.~\ref{fig:DPEnergyLossLZ}, in which we considered a simulation with $v_g = 0.3$. The computation is performed both in the mass (blue curve) and interaction (orange curve) bases and, in order to better visualize the discrepancy, we plot the relative difference in the lower panel. We can clearly see the decrease in energy happening when the wave packet of the dark photon crosses the resonant point, undergoing the conversion process. In the early stages of the simulation the energy increases due to the fact that the dark photon wave packet is entering in the interval over which the energy is computed. Furthermore, well after the conversion process, the energy displays a second small decrease in a step-like behavior (inset of Fig.~\ref{fig:DPEnergyLossLZ}), which comes from the fact that a small component of the dark photon is reflected by the plasma barrier and eventually passes the position $z_0$, exiting the integration domain.

\begin{figure}
    \includegraphics[width = \columnwidth]{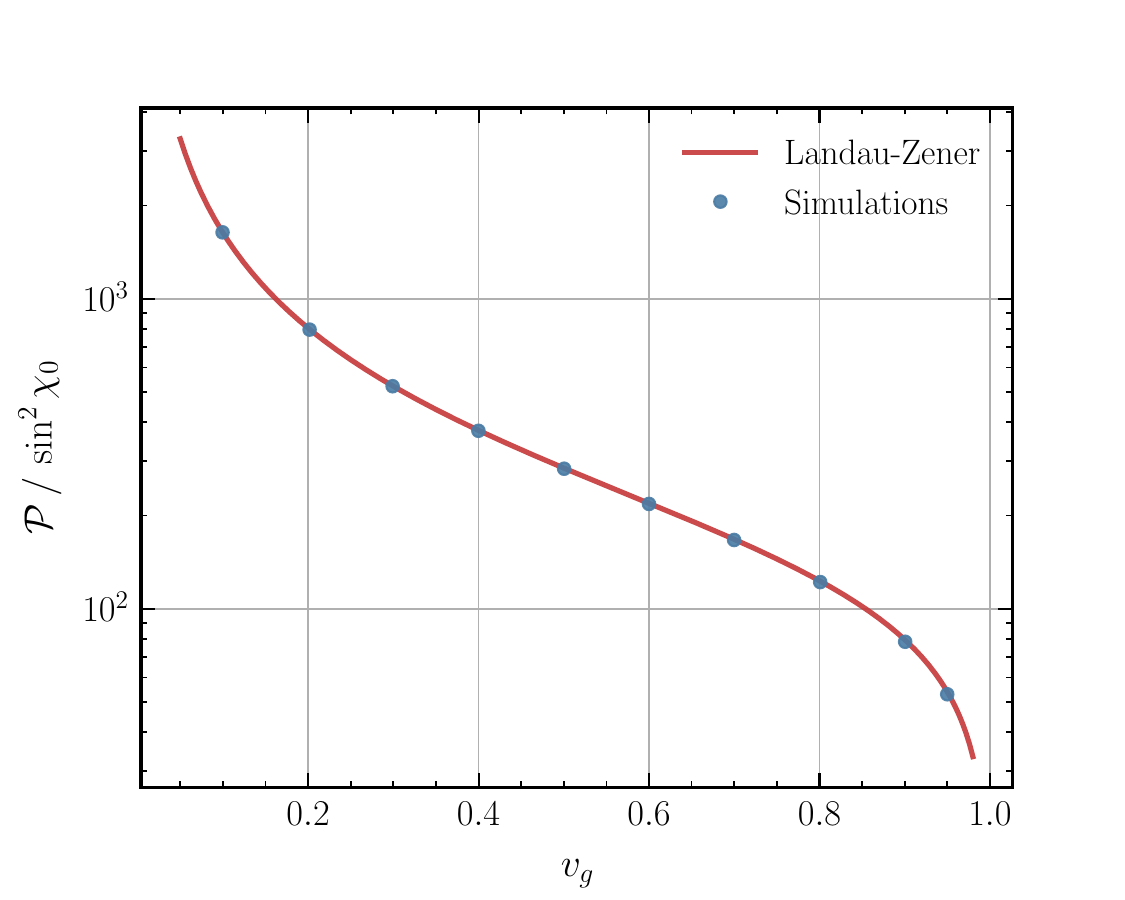}
    \caption{
    Comparison between the conversion probability computed from the LZ formula \eqref{eq:LZformula} (red solid curve), and the probability obtained from our numerical simulations (blue points) in the set with varying group velocity of the dark photon wave packet. In the regime we analyzed using this set of simulations, the agreement is excellent.
    \label{fig:LZProbabilityComparison}
    }
\end{figure}

We produced similar plots for each simulation in the set, and from them we extracted the energy of the dark photon before and after the conversion process. We then computed the probability $\PP$ using Eq.~\eqref{eq:LZSimulationProbability} and we compared the result against Eq.~\eqref{eq:LZformula}. We performed these operations only in the interaction basis, as it is the one in which the Landau-Zener formula holds. The results are shown in Fig.~\ref{fig:LZProbabilityComparison}, where we can clearly see an excellent agreement of our numerical results with the estimates from Eq.~\eqref{eq:LZformula}.

\subsection{Breakdown of the LZ formula with multiple level crossings} \label{sec:results:MLC}

\begin{figure}
    \includegraphics[width=\columnwidth]{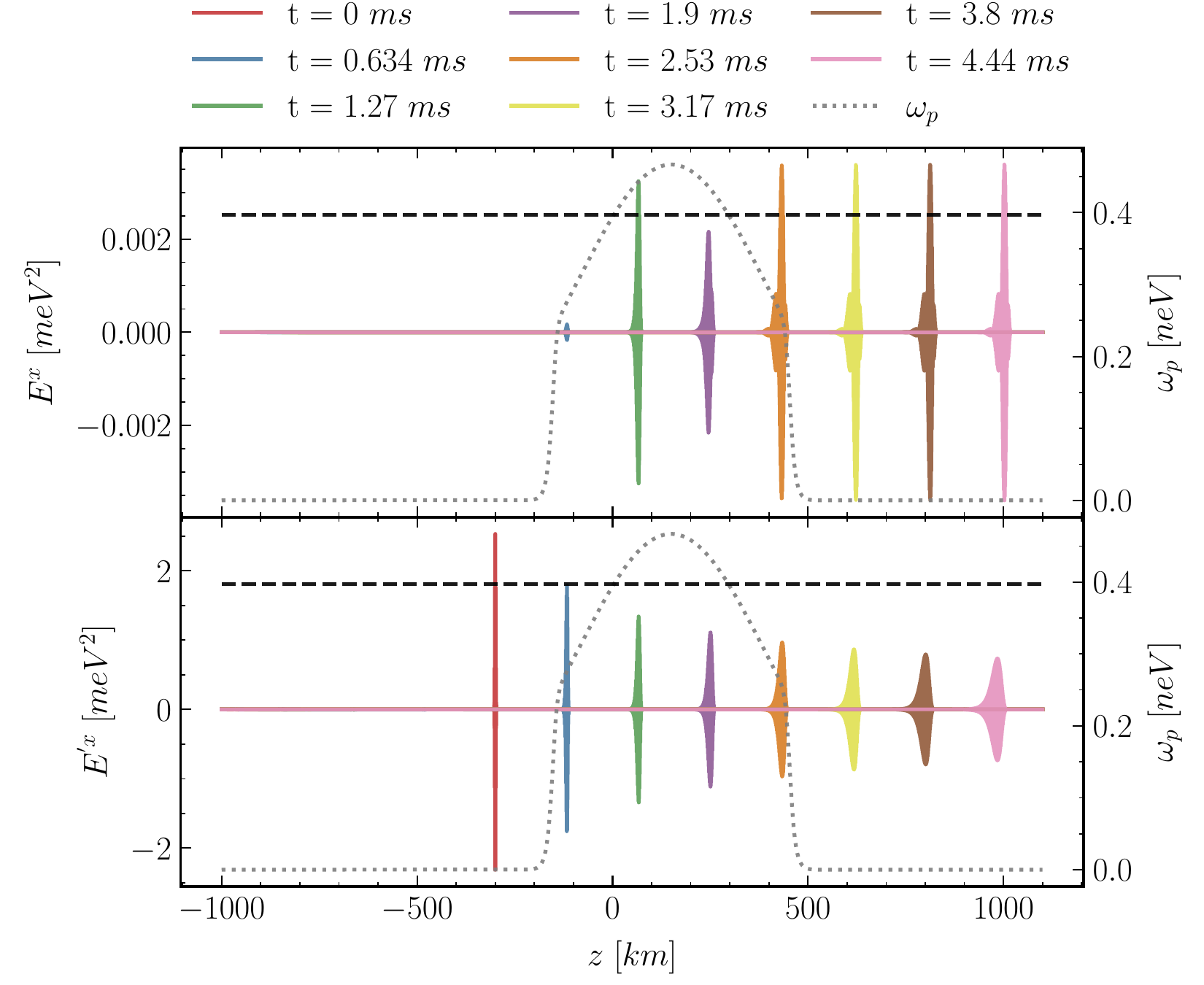}
    \caption{Snapshots of a typical simulation in the set for testing the breakdown of LZ in the presence of multiple level crossings. Specifically, this figure shows the case $\mu = 0.85 \, \omega_{p, \rm max}$ using the same conventions as in Fig.~\ref{fig:LZsnapshots}. The fields are represented in the mass eingenbasis, which is the one used in the numerical integration. As we can see two wavepackets of ordinary electromagnetic fields are produced when the dark photon crosses the resonant points. Here, since the frequency of the photon is larger than $\omega_{p, \rm max}$, the photon is able to propagate through plasma, but undergoing dispersion and travelling at a smaller group velocity than the dark photon. This results in the second photon being produced in front of the first one, and being more localized in space compared to the first one.
    \label{fig:MLCSnapshots}
    }
\end{figure}

In the second set of simulations we studied the case in which the resonance occurs in the vicinity of a local maximum of the plasma frequency, with the aim of checking, with a time-domain analysis, the results recently found in Ref.~\cite{Brahma:2023zcw} using frequency-domain computations. To reproduce a plasma profile with a local maximum, instead of a plateau, we chose values of $z_1$ and $z_2$ close to each other, and we decreased $W_{1,2}$ to make the peak less sharp. However, with small values of $W_{1,2}$ the distance between the position of the peak and the region where the plasma density reaches its background value is considerably large. 
Since we wish to place the initial wave packet of the dark photon in this latter region, then the dark photon would take a long time to reach the resonant points, resulting in large computational costs. We decided therefore to modify the plasma profile~\eqref{eq:InitialPlasmaDensity} by shortening its tails outside the resonant points, in order to be able to initialize the simulation with the dark photon located closer to the center of the barrier. The corresponding plasma frequency profile is presented in Fig.~\ref{fig:MLCSnapshots} (gray dotted line in both panels). We achieved this by multiplying the original barrier of height $n_{\rm max} - n_{\rm bkg}$ by an analogous barrier of height 1, resulting in:
\begin{align}
    n_0 &= n_{\rm bkg} + (n_{\rm max} - n_{\rm bkg}) \notag \\
                   &\times\Bigl[\sigma(z; W_1, z_1) + \sigma(z; -W_2, z_2) - 1\Bigr] \notag \\
                   &\times\Bigl[\sigma(z; W_s, z_c - \Delta z_s) + \sigma(z; -W_s, z_c + \Delta z_s) - 1\Bigr],
    \label{eq:InitialPlasmaDensityShortenedTails}
\end{align}
where $z_c = (z_1 + z_2)/2$ is the center of both barriers, $\Delta z_s$ is the distance between the boundaries of the new barrier and $z_c$, and $W_s$ controls their slope. In this way, if $\Delta z_s$ and $W_s$ are sufficiently large, the new term will dampen the original barrier profile for $z \lesssim z_c - \Delta z_s$ and $z \gtrsim z_c + \Delta z_s$, while preserving its shape in the central region. 
As for the numeric values of the parameters that we used, we set $n_{\rm bkg} = 10^{-6} \, \cm^{-3}$ and $n_{\rm max} = 250 \, \cm^{-3}$, resulting in plasma frequencies $\omega_{p, \rm bkg} = 3.7045 \times 10^{-14} \, \eV$ and $\omega_{p, \rm max} = 4.6681 \times 10^{-10} \, \eV$ for the background and the top of the barrier, respectively; we placed the boundaries at $z_1 = 0 \, \km$ and $z_2 = 300 \, \km$, so that $z_c = 150 \, \km$, and we set their steepness parameters to $W_{1,2} = 0.01 \, \km^{-1}$; lastly, we set $\Delta z_s = 300 \, \km$, and $W_a = 0.2 \, \km^{-1}$.

In this set of simulations we varied the dark photon mass $\mu$ in the range $0.76 \, \omega_{p, \rm max} \le \mu \le 0.9999 \, \omega_{p, \rm max}$, affecting the position of the resonant points and pushing them considerably close to the peak of the barrier. The frequency $\omega$ appearing in Eq.~\eqref{eq:EMInitial} was set to $\omega = 1.5 \times 10^{-9} \eV$, that results in a group velocity $v_g = 0.95 \div 0.97$, depending on the value of $\mu$. The width of the wave packet was set to $\sigma = 1 \, \km$, its initial position to $z_0 = -300 \, \km$, and its amplitude to $A_E = m_e \omega / (1000 e)$. 

Lastly, our numerical grid extends in $-1000 \, \km \le z \le 1100 \, \km$, and the grid step is $\Delta z = 10^8 \, \eV^{-1} = 19.7 \, \m$, consistently with the criterion indicated in Sec.~\ref{sec:simparams}.

In Fig.~\ref{fig:MLCSnapshots} we show some snapshots of the evolution in the case $\mu = 0.85 \, \omega_{p, \rm max}$, using the same conventions as in Fig.~\ref{fig:LZsnapshots} and expressing the fields in the mass eingenbasis. In the first stages of the simulations the wave packet of the dark photon field travels towards the plasma barrier, undergoing a dispersion process. When it crosses the first resonant point, a photon is produced which, unlike the case discussed in Sec.~\ref{sec:results:LZ}, is able to propagate through the barrier. In this stage the photon has an effective mass $\omega_p  \approx \omega_{p, \rm max}$, and undergoes a dispersion phenomenon. When the dark photon crosses the second resonance, another photon is produced in front of the first one, due to the fact that $\omega_{p, \rm max} > \mu$, and inside the barrier the group velocity of the dark photon is larger than the group velocity of the ordinary photon. Finally, the two photons propagate undergoing a negligible dispersion since their effective mass is $\omega_{p, \rm bkg}$.

\begin{figure}
    \includegraphics[width=\columnwidth]{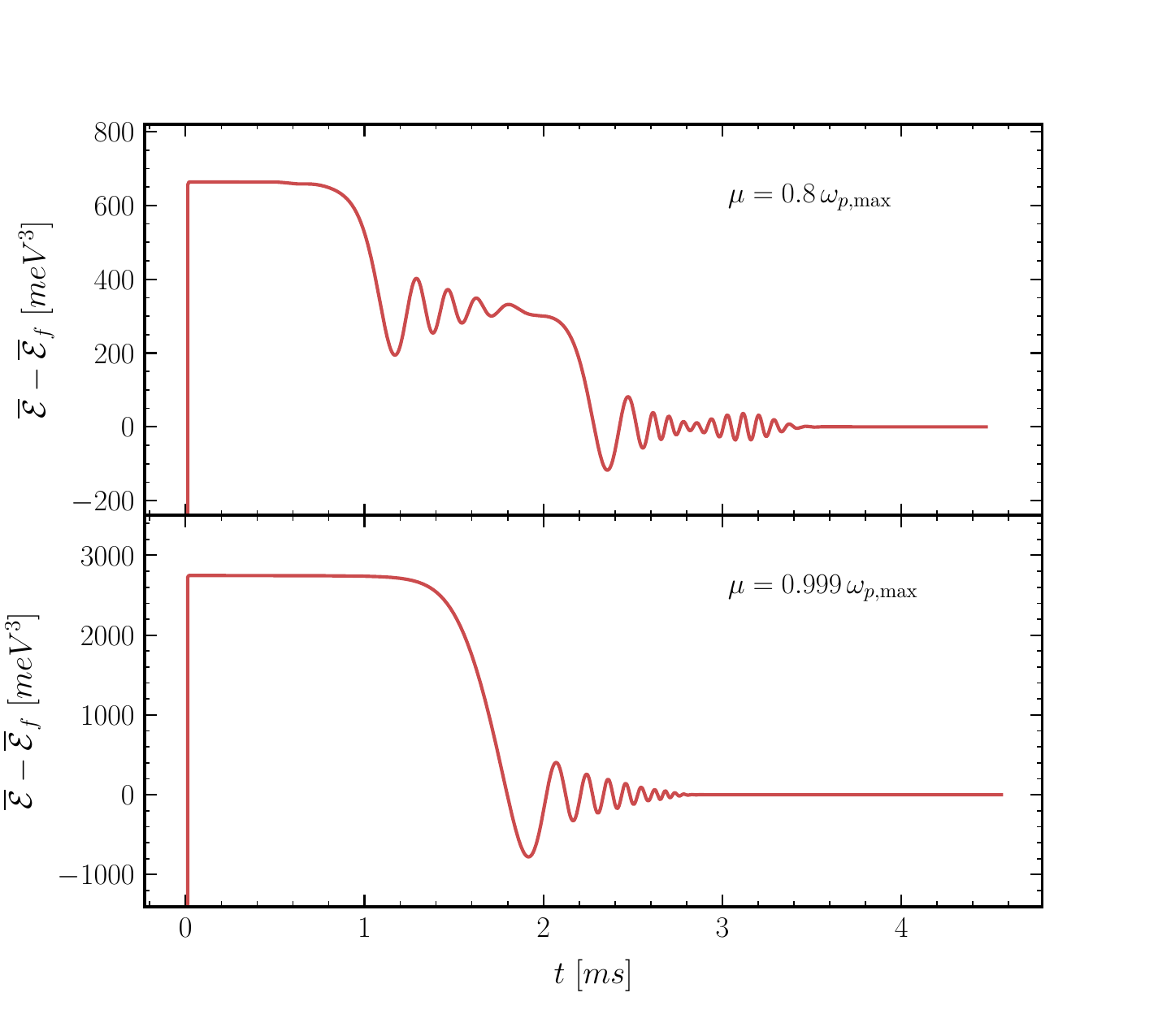}
    \caption{Representative plot of the behavior of the energy per unit surface of the dark photon in the interaction basis, for the set of simulations aimed at testing the violation of the LZ law in presence of multiple level crossings. The upper panel shows the case in which the two resonant points are well separated, while the lower panel the case in which they are close to each other. As we can see in the former case we can clearly recognize both the conversions, while in the latter case they visually appear as a single-step loss in energy.
    \label{fig:MLCEnergy}
    }
\end{figure}

As a first step towards the computation of the conversion probability, we computed the energy per unit surface of the dark photon in the interaction basis using Eq.~\eqref{eq:DPEnergy}. Interestingly, when $\mu$ is sufficiently small the resonant points are distant enough that it is possible to identify the two transitions separately. On the contrary, when the resonant points are close, only a single energy loss step appears. This is visible in Fig.~\ref{fig:MLCEnergy}, where we show the behavior of the energy per unit surface, $\overline \EE$, in the two cases discussed.

As in the previous set of simulations, the conversion probability can be extracted from the value of $\EE$ before and after the interaction with the plasma barrier. However, since here there are two resonant points, we must generalize Eq.~\eqref{eq:DPEnergyLossLZ}, which only accounts for a single conversion. In this case the expression of $\EE_f$ takes contributions from two different processes. In the first the component of the dark photon emerging from the first resonant point undergoes a further energy loss due to the second conversion, yielding
\begin{equation}
    \EE^{\rm (1)}_f = (1 - \PP_1)(1 - \PP_2) \EE_i = \EE_i -  (\PP_1+\PP_2) \EE_i + \OO{\PP^2}, 
    \label{eq:DPEnergyLossMLCTerm1}
\end{equation}
where $\PP_{1,2}$ are the transition probabilities at the two resonant points.
In the second process, instead, the ordinary photon produced from the first resonant point, converts part of its energy to the dark photon field when crossing the second resonant point, and contributes with a term $\EE^{\rm (2)}_f = \OO{\PP^2}$. Neglecting terms of order $\PP^2$ we can then write
\begin{equation}
    \PP_{\rm tot} = \PP_1+\PP_2= \frac{\EE_i - \EE_f}{\EE_i}.
    \label{eq:MLCSimulationProbability}
\end{equation}

\begin{figure}
    \centering
    \includegraphics[width=\columnwidth]{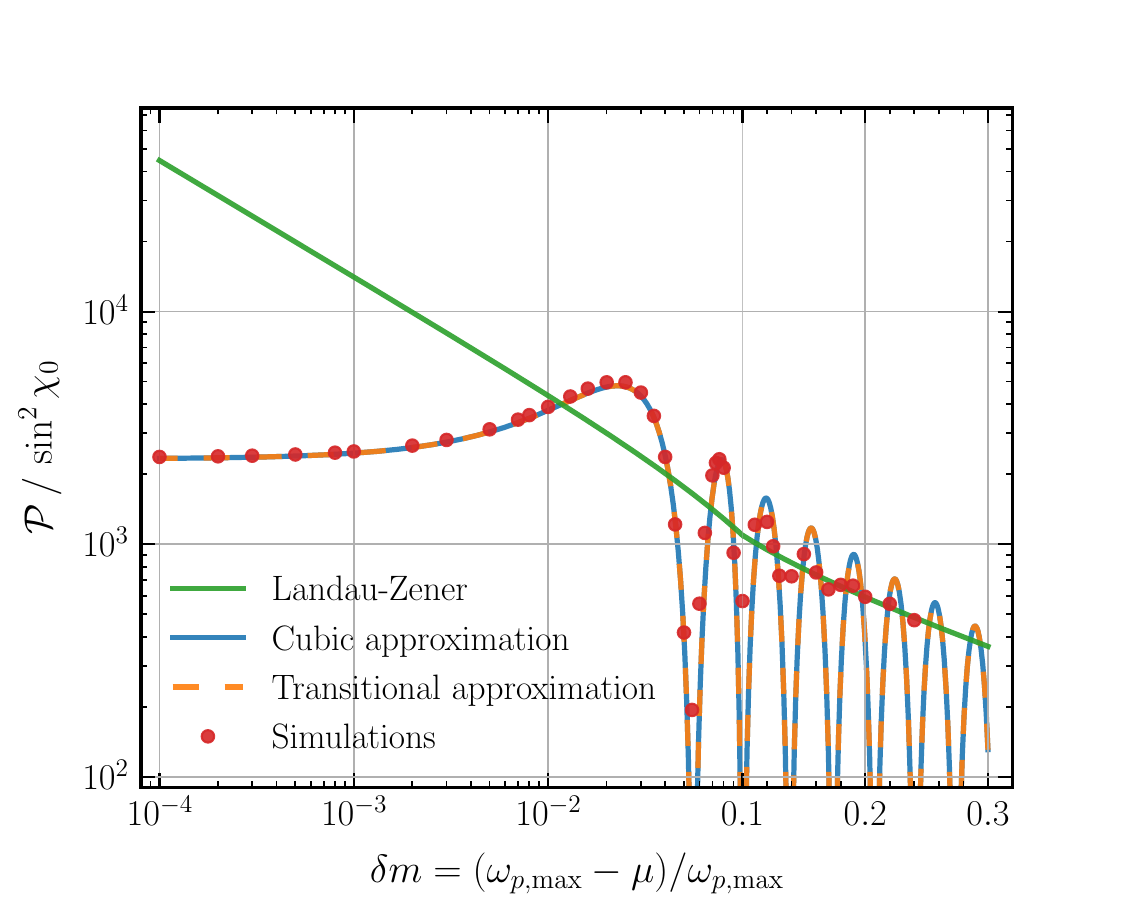}
    \caption{Conversion probability for the set of simulations in which the dark photon crosses two resonant points. Green, blue and orange lines denote the conversion probability estimated with the LZ formula \eqref{eq:LZformula}, cubic approximation \eqref{eq:CubicApproximation}, and transitional Airy approximation \eqref{eq:TransitionalAiryApproximation}, respectively. Red points instead denote the probabilities extracted from our simulations. As we can see our results are in excellent agreement with those obtained in Ref.~\cite{Brahma:2023zcw}, and in the case in which the dark photon mass is close to the plasma density at the peak, the conversion probability is more than an order of magnitude smaller than the prediction by the LZ formula. On the contrary,  for large values of $\delta m$ the probability approaches the LZ estimate.
    \label{fig:MLCProbabilityComparison}
    }
\end{figure}

In Fig.~\ref{fig:MLCProbabilityComparison} we compare the conversion probabilities obtained from our simulations (in red) with the predictions obtained using the LZ formula (green), the cubic approximation (blue) and the transitional approximation (orange). We can immediately see that when $\mu$ is close to $\omega_{p, \rm max}$, and the two resonant points are close to each other, the LZ formula tends to overestimate the conversion probability by more than a order of magnitude. Our result show instead a remarkable agreement with analysis performed in~\cite{Brahma:2023zcw}. The cubic and transitional approximations, which should not break down near coalescing saddle points, are in excellent agreement with the results of our simulations, and correctly estimate the probability in the coalescing limit. As $\delta m$ increases, phase effects between the two resonant points cause oscillations to appear. This behavior is captured by our simulations, and when the separation between the resonant points becomes large enough, the conversion probability approaches the estimate obtained with the LZ formula.

\subsection{Estimate of the conversion probability beyond the WKB approximation} \label{sec:results:vW}

Finally, we performed a set of simulations in which we varied the slope of the plasma frequency at the resonant point, with the aim to test the validity of the LZ formula as the slowly-varying background plasma approximation ceases to be valid. We considered the plasma profile as in the set discussed in Sec.~\ref{sec:results:LZ}, with the difference that here we varied the parameter $W$ in the range $0.01 \, \km^{-1} \le W \le 3 \, \km^{-1}$. We considered a dark photon with mass $\mu = 10^{-10} \, \eV$, and we set the parameter $\omega$ in Eq.~\eqref{eq:EMInitial} to $\omega = 1.4 \times 10^{-10} \, \eV$, so that the group velocity is $v_g = 0.7$. We set the width of the wave packet to $\sigma = 30 \, \km$, and its position to $z_0 = -3500 \, \km$. The numerical grid extended from $z_{-\infty} = -10000 \, \km$ to $z_{+\infty} = 4000 \, \km$, with a grid step $\Delta z = 1.12 \times 10^9 \, \eV^{-1} \approx 221 \, \m$. While this is in agreement with the criterion introduced in Sec.~\ref{sec:simparams}, for the two simulations in which the plasma barrier is steeper, i.e.\ the ones with $W = 2.3 \, \km^{-1}$ and $W = 3 \, \km^{-1}$, we halved the spatial step in order to resolve properly the boundary of the barrier. Given the similarity of the setup with the one described in Sec.~\ref{sec:results:LZ}, we did not repeat the simulation with $W = 0.01 \, \km$ but we used the simulation with $v_g = 0.7$ we performed as part of that set.

\begin{figure}
    \includegraphics[width=\columnwidth]{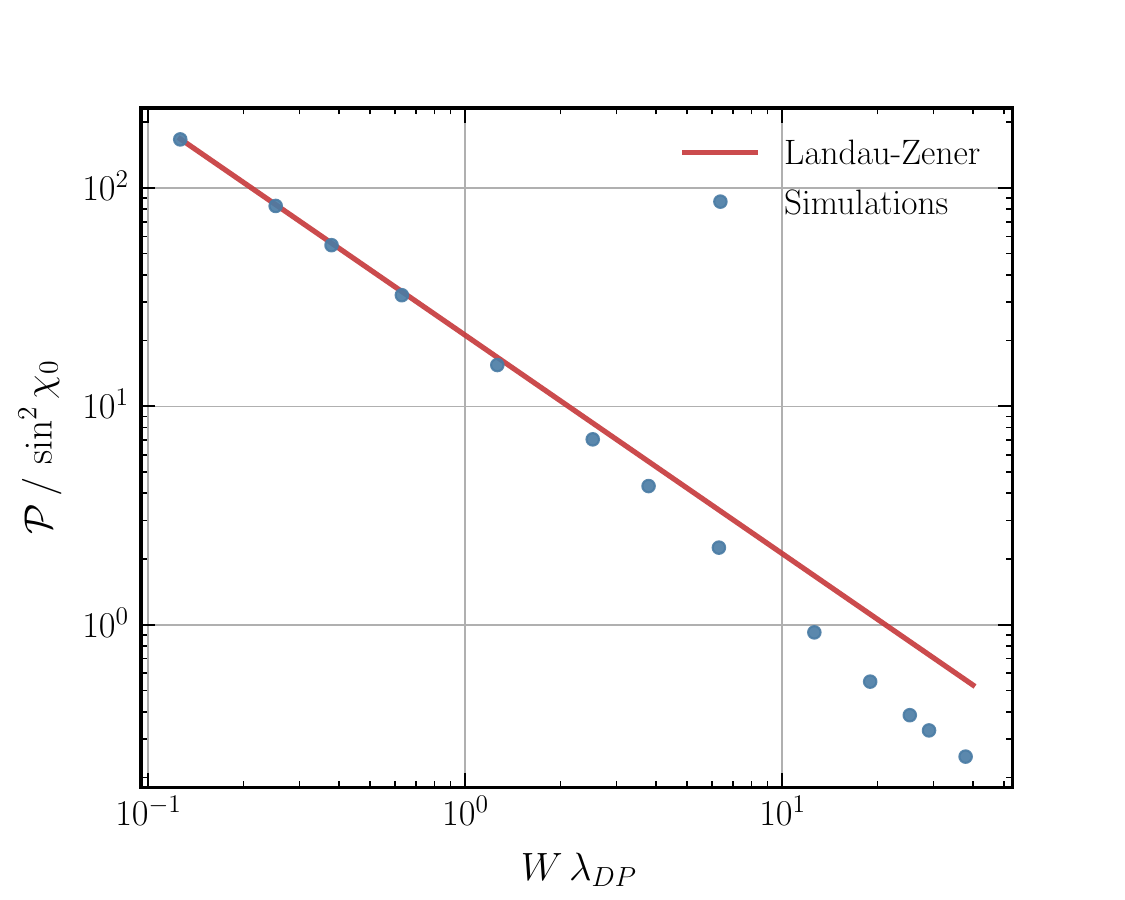}
    \caption{Conversion probability for the setup described in Sec.~\ref{sec:results:vW}, as a function of the ratio between the dark photon wavelength, $\lambda_{\rm DP}$, and the scale over which the plasma frequency varies, $1/W$. The red curve represents the prediction by the LZ formula \eqref{eq:LZformula}, and the blue points represent the probability extracted from the numerical simulations. As we can see, when the steepness of the barrier increases, the LZ formula starts overestimating the conversion probability, due to the fact that the approximation of a slowly-varying plasma gradually loses validity. The reduction we find here seems in agreement with Ref.~\cite{Beadle:2024jlr}, where the authors studied in the frequency domain the resonant conversion process in the Earth ionosphere, where the plasma is not slowly-varying and LZ was shown to fail.
    \label{fig:vWProbabilityComparison}
    }
\end{figure}

Since here the dark photon undergoes a single conversion process, we extracted the probability $\PP$ using Eq.~\eqref{eq:LZSimulationProbability}. The results are shown in Fig.~\ref{fig:vWProbabilityComparison}, together with the estimate obtained from the LZ formula. As we can see when $W$ assumes large values, and the length scale over which the plasma density varies becomes comparable to the dark photon wavelength, the LZ prediction tends to overestimate the conversion probability. This seems to confirm the recent results presented in Ref.~\cite{Beadle:2024jlr}, where the authors studied (in the frequency domain) the resonant conversion process in the Earth ionosphere, a prototypical case where the slowly-varying plasma density approximation is not a good one, and LZ fails.

\section{Conclusion and outlook}

In this work, we developed a versatile 1+1 code, designed to perform non-linear simulations of systems involving photons coupled to ultralight bosons in the presence of plasma. This framework provides a powerful tool for exploring resonance conversions, particularly in cases where the LZ approximation is known to break down, such as in scenarios with multiple level crossings or where the de Broglie wavelength of the new boson is comparable to the spatial variations in the plasma. As a first application of this code, we have studied the resonance conversion in non-turbulent plasma in 1+1 dimensions, providing a crucial step forward in simulating these processes in realistic astrophysical and cosmological contexts. Our simulations highlight the limitations of the original LZ approximation while showing a remarkable agreement with the extensions recently computed in Ref.~\cite{Brahma:2023zcw} in the frequency domain. Furthermore, our results do not rely on the WKB approximation, so are valid for any frequency and any plasma profile. 

This initial study lays the groundwork for more complex investigations into resonance conversion phenomena in a wide range of environments. 
Future work will extend the code’s applicability to 3+1 scenarios, possibly involving both turbulent and anisotropic plasmas. In the former case, small-scale fluctuations play a significant role and can further impact the conversion processes, while in the latter photons and light bosons move along different worldlines, rendering standard analytical approaches challenging~\cite{Witte:2021arp, McDonald:2023ohd}.
Our framework can thus be adapted to explore a variety of astrophysical and cosmological scenarios, where light bosons interact with photons in plasma-rich environments. Although we have focused on dark photons in this study, the extension to axions is straightforward. In this context, future directions include exploring dark photon and axion dark matter conversion in the Earth's ionosphere~\cite{Beadle:2024jlr}, the solar corona~\cite{An:2023mvf, An:2023wij}, and neutron star magnetospheres~\cite{Gines:2024ekm}.

The flexible nature of our code also allows for applications in completely different setups where real-time computations are crucial, such as in direct detection experiments for light bosons. In such cases, particular attention must be paid to boundary conditions and symmetries~\cite{Jeong:2023bqb}.

Our ultimate goal is to make the code public and user-friendly after completing additional studies to validate its broader applicability.

\begin{acknowledgments}
We thank Samuel Witte, Sebastian Ellis, Jamie Mcdonald and Nirmalya Brahma for useful comments on the draft.
E.C. acknowledges financial support provided under the European Union’s H2020 ERC Advanced Grant “Black holes: gravitational engines of discovery” grant agreement no. Gravitas–101052587. Views and opinions expressed are however those of the author only and do not necessarily reflect those of the European Union or the European Research Council. Neither the European Union nor the granting authority can be held responsible for them. E.C. acknowledges support from the Villum Investigator program supported by the VILLUM Foundation (grant no. VIL37766) and the DNRF Chair program (grant no. DNRF162) by the Danish National Research Foundation. This project has received funding from the European Union's Horizon 2020 research and innovation programme under the Marie Sklodowska-Curie grant agreement No 101007855 and No 101131233.
This work is partially supported by the MUR PRIN Grant 2020KR4KN2 ``String Theory as a bridge between Gauge Theories and Quantum Gravity'', by the FARE programme (GW-NEXT, CUP:~B84I20000100001), and by the INFN TEONGRAV initiative.
The numerical simulations have been performed at the Vera cluster supported by the Italian Ministry for Research and by Sapienza University of Rome.
\end{acknowledgments}

\appendix

\section{Convergence of the code} \label{app:convergence}

To monitor the convergence of our code we studied the scaling with resolution of the violation of Eqs.~\eqref{eq:Gauss},~\eqref{eq:DarkGauss}, and~\eqref{eq:PlasmaConstraint}, which read
\begin{align}
    CV_{\rm photon} &= \partial_i E^i - e \nel - \rho_{\rm (ions)},  \label{eq:CVphoton}\\
    CV_{\rm dark~photon} &= \partial_i \dE^i - \snc (e \nel + \rho_{\rm (ions)}) + \mu^2 \dvarphi, \label{eq:CVdarkphoton} \\
    CV_{\rm plasma} &= \sqrt{\Gamma^2 (1 - \UU_i \UU^i)} - 1. \label{eq:CVplasma}
\end{align}

\begin{figure}
    \centering
    \includegraphics[width=\columnwidth]{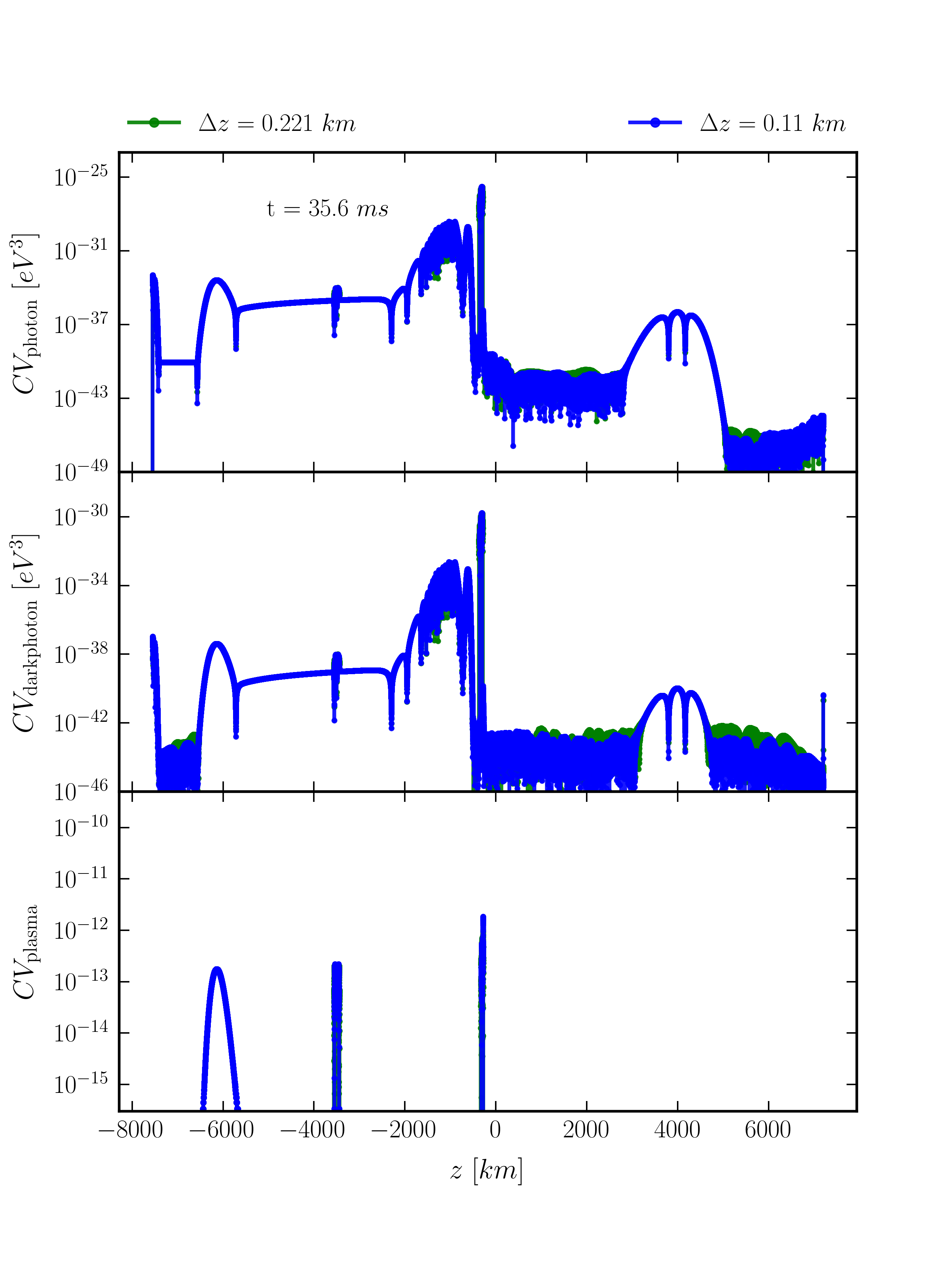}
    \caption{Constraint violations for the simulation with initial group velocity of the wave packet of the dark photon $v_g = 0.7$, at time $t = 35.6 \, \ms$ using two different resolutions. As we can see in this regime the constraint violations do not satisfy the scaling behavior expected from the characteristics of the evolution code.
    \label{fig:LinearRegimeConvergence}
    }
\end{figure}

We ran a first convergence test on the simulation with initial group velocity of the dark photon wave packet $v_g = 0.7$, from the set discussed in Sec.~\ref{sec:results:LZ}. In particular we repeated such simulation halving the grid step $\Delta z$, together with the time step $\Delta t$ in order to keep the CFL factor constant. We show in Fig.~\ref{fig:LinearRegimeConvergence} the constraint violations at time $t = 35.6 \, \ms$, close to the end of the simulation. As we can see the constraint violation is extremely small, but it does not scale with resolution.

\begin{figure}
    \includegraphics[width=\columnwidth]{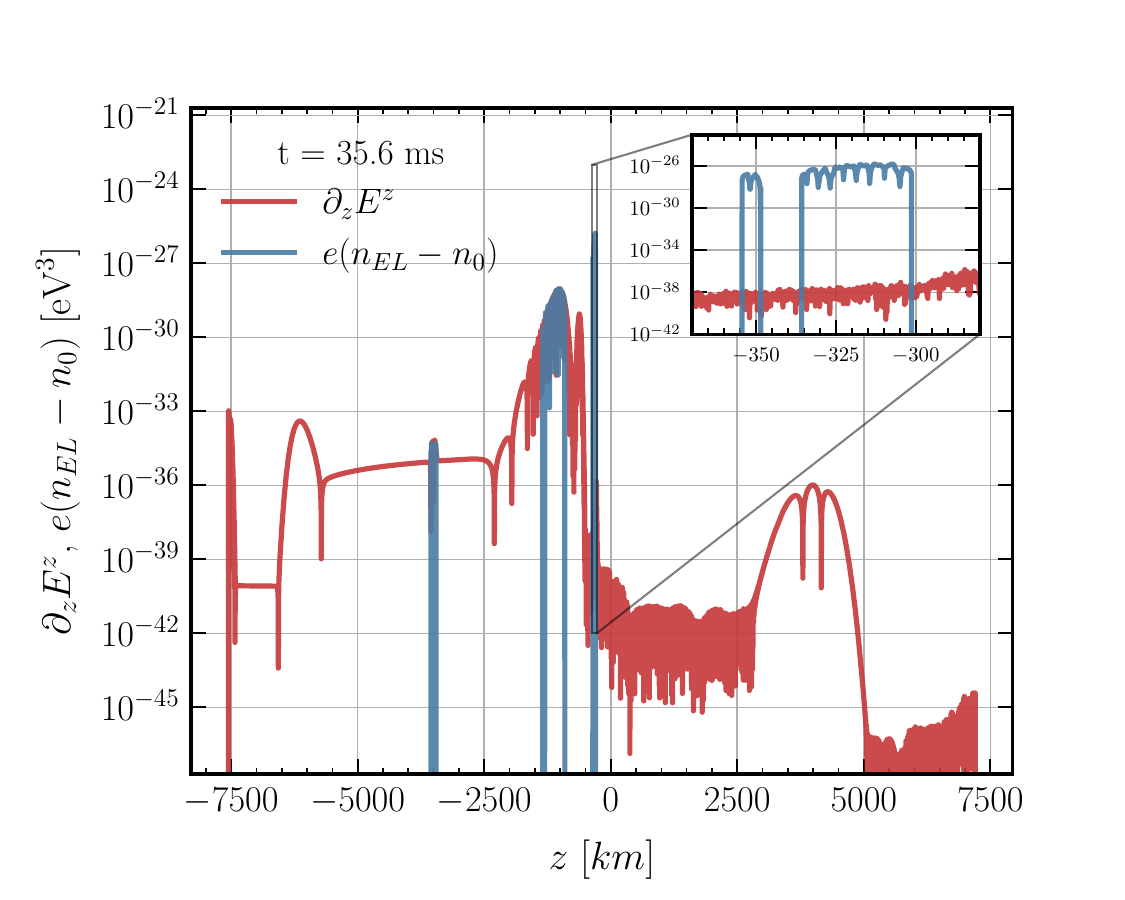}
    \caption{Behavior of the terms composing the violation of the Gauss law for the photon, in the profile shown in the upper panel of Fig.~\ref{fig:LinearRegimeConvergence}. As we can see the density term $e(\nel - n_0)$ is non-vanishing only in limited intervals, meaning that in the rest of the spatial domain, at $t = 35.6 \, \ms$ $\nel$ assumes its initial value.
    \label{fig:CVPhotonTermsLinear}
    }
\end{figure}

To better investigate the motivations behind this behavior, let us consider the violation of the Gauss law for the photon, Eq.~\eqref{eq:CVphoton}, which is composed by two terms that are subtracted: $\partial_z E^z$ and $\rho = e \nel + \rho_{\rm (ions)} = e (\nel - n_0)$. As we can see from Fig.~\ref{fig:CVPhotonTermsLinear}, in which we plotted them separately, the charge density is non-zero only in delimited regions, outside which $\nel$ still assume its initial value, $n_{\rm EL} = n_0$, and has likely not evolved.

\begin{figure}
    \includegraphics[width=\columnwidth]{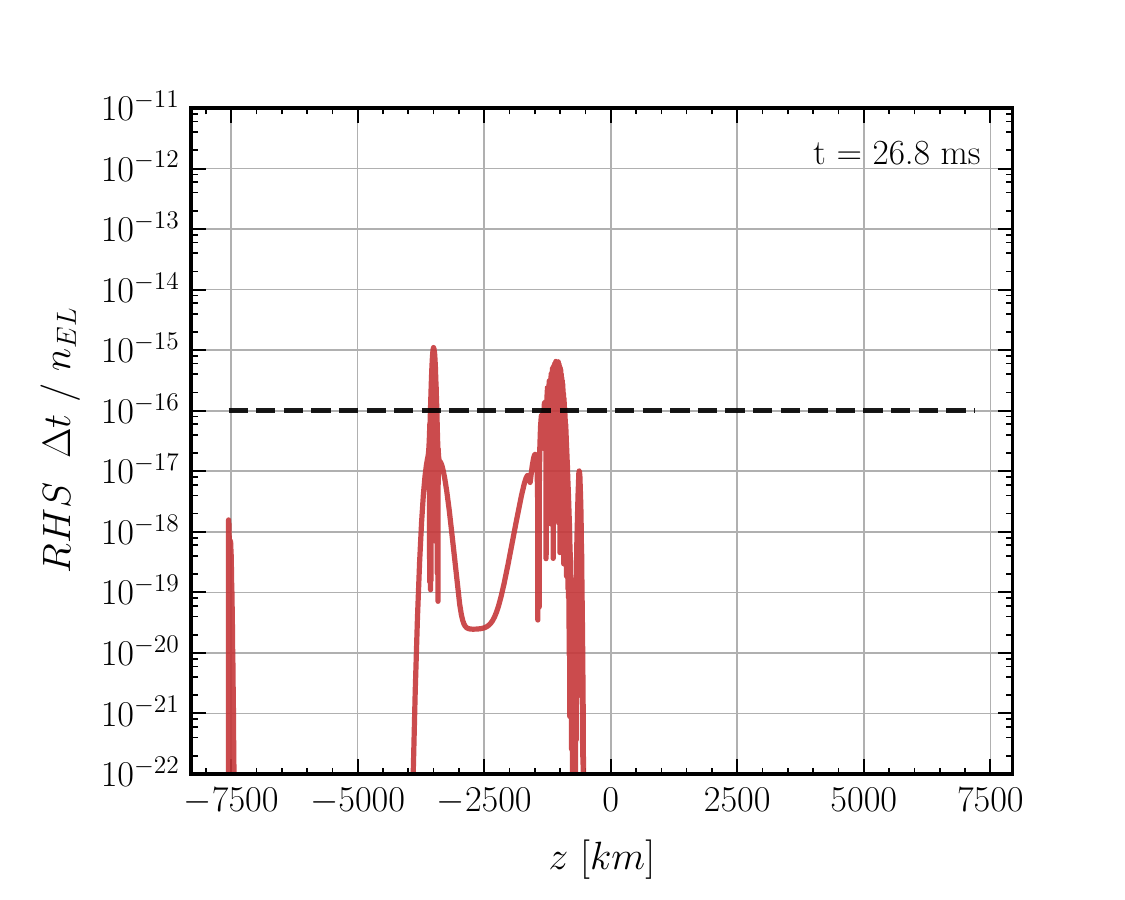}
    \caption{Relative increment of $\nel$ for the simulation with $v_g = 0.7$, at $t = 26.8 \, \ms$. The horizontal dashed line marks the value $10^{-16}$, which is the order of magnitude of machine precision error. This plot shows that the evolution of $\nel$ mostly falls below machine precision, and can be resolved by our code only in limited intervals.
    \label{fig:LinearNelEvolutionTerm}
    }
\end{figure}

In our integration algorithm time evolution is performed with the Runge-Kutta method. At each integration step the profile of $\nel$ is updated by adding a term which, for our purposes here, we can roughly estimate as ${\rm RHS} \, \Delta t$, where $\Delta t$ is the time step, and RHS is the right hand side of Eq.~\eqref{eq:NELEvol}, i.e.
\begin{equation}
    {\rm RHS} = - \UU^i \partial_i \nel - \nel \partial_i \UU^i.
\end{equation}
In Fig.~\ref{fig:LinearNelEvolutionTerm} we show the profile of the relative increment of $\nel$ at $t = 26.8 \, \ms$, earlier than the time at which the profiles in Fig.~\ref{fig:LinearRegimeConvergence} are extracted. The horizontal dashed line marks the order of magnitude of machine precision, which is smaller than ${\rm RHS} \, \Delta t \, / \, \nel$ only in intervals that roughly corresponds to two where $\rho \neq 0$ at $t = 35.6 \, \ms$. By performing a similar plot at $t = 15.7 \, \ms$ we obtained ${\rm RHS} \, \Delta t \, / \, \nel > 10^{-16}$ for $-370 \, \km \lesssim z \lesssim -290 \, \km$. This scenario seems to suggest that the evolution of $n_{\rm EL}$ mostly falls below machine precision, and cannot be resolved by the code, leading to the scaling behavior observed in the top panel of Fig.~\ref{fig:LinearRegimeConvergence}.

To assess the reliability of our code we therefore decided to simulate a scenario in which the backreaction effects on plasma are larger. We considered the same setup, but starting with a wave packet of the ordinary photon, for which the mass appearing in Eq.~\eqref{eq:EMInitial} is given by the plasma frequency of the background. We set the initial amplitude to $A_E = m_e \omega / (10 e)$, so that we are closer to the nonlinear regime of interaction between the electromagnetic field and plasma. These simulations crashed at $t \approx 26 \, \ms$ with the plasma density being subject to high frequency noise at the base of the barrier. Since here we are only interested in evaluating the convergence of our integration algorithm, and the nonlinear regime of interaction is outside the scope of this paper, presenting its own difficulties in the application of the fluid approximation (cf. Ref.~\cite{Cannizzaro:2023ltu}), we did not investigate this problem further.

\begin{figure}
    \includegraphics[width=\columnwidth]{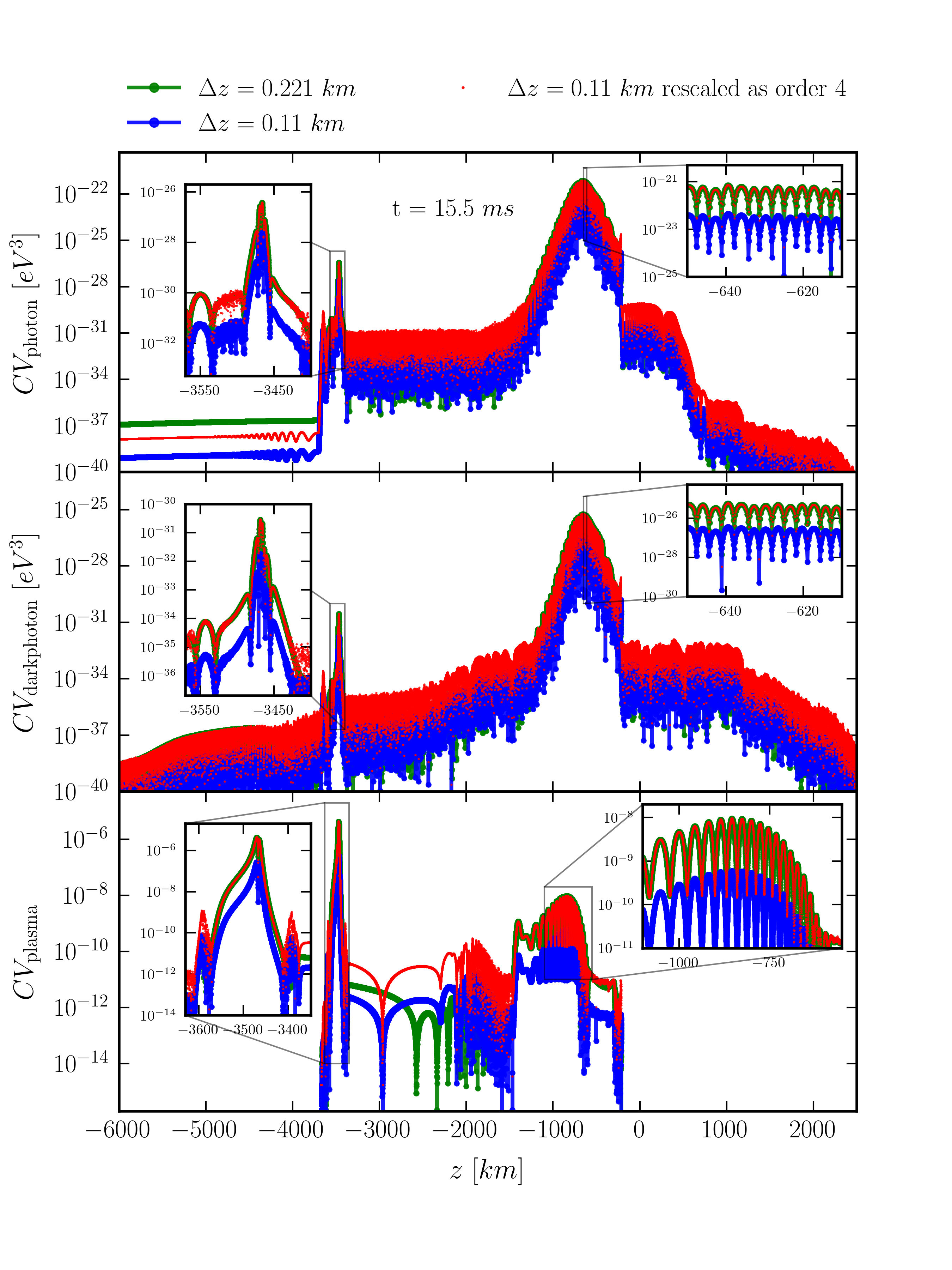}
    \caption{Scaling of the constraint violations in a simulation of the scattering of a wave packet of the photon on a barrier of plasma, in a regime where nonlinearities are relevant. The constraint violations for the simulations with higher and lower resolution are shown in blue and green, respectively, while the red dots represent the constraint violations for the simulation with higher resolution rescaled by a factor corresponding to fourth-order convergence. The data are extracted at time $t = 15.5 \, \ms$. The constraint violations possess two peaks where the expected scaling behavior is found, while in the rest of the domain convergence is lost, and in some regions it is dominated by noise. However, in such regions the constraint violations is smaller than on the peaks by $1 \div 4$ orders of magnitude.
    \label{fig:NonlinearLZConvergence}
    }
\end{figure}

In Fig.~\ref{fig:NonlinearLZConvergence} we show the scaling behavior of the constraint violations at $t = 15.5 \, \ms$, sufficiently earlier than the crash. In blue and green we denote the constraint violations for the simulation with higher and lower resolution, respectively, whereas in red we show the constraint violations in the simulation with higher resolution rescaled by a factor $\bigl( \Delta z_{\rm coarse} / \Delta z_{\rm fine} \bigr)^4$, corresponding to a convergence of order 4. This is the scaling behavior that we expected, since in our numerical setup we compute spatial derivatives at fourth order of accuracy and we perform time integration with a sixth order accurate algorithm. As we can see all the constraint violations have two peaks where they satisfy the expected scaling behavior considerably well. Outside these two peaks the convergence is lost, and in some regions it is dominated by noise, but it generally assumes values that are $1 \div 4$ orders of magnitude smaller than on the peaks. 

\begin{figure}
    \centering
    \includegraphics[width=\columnwidth]{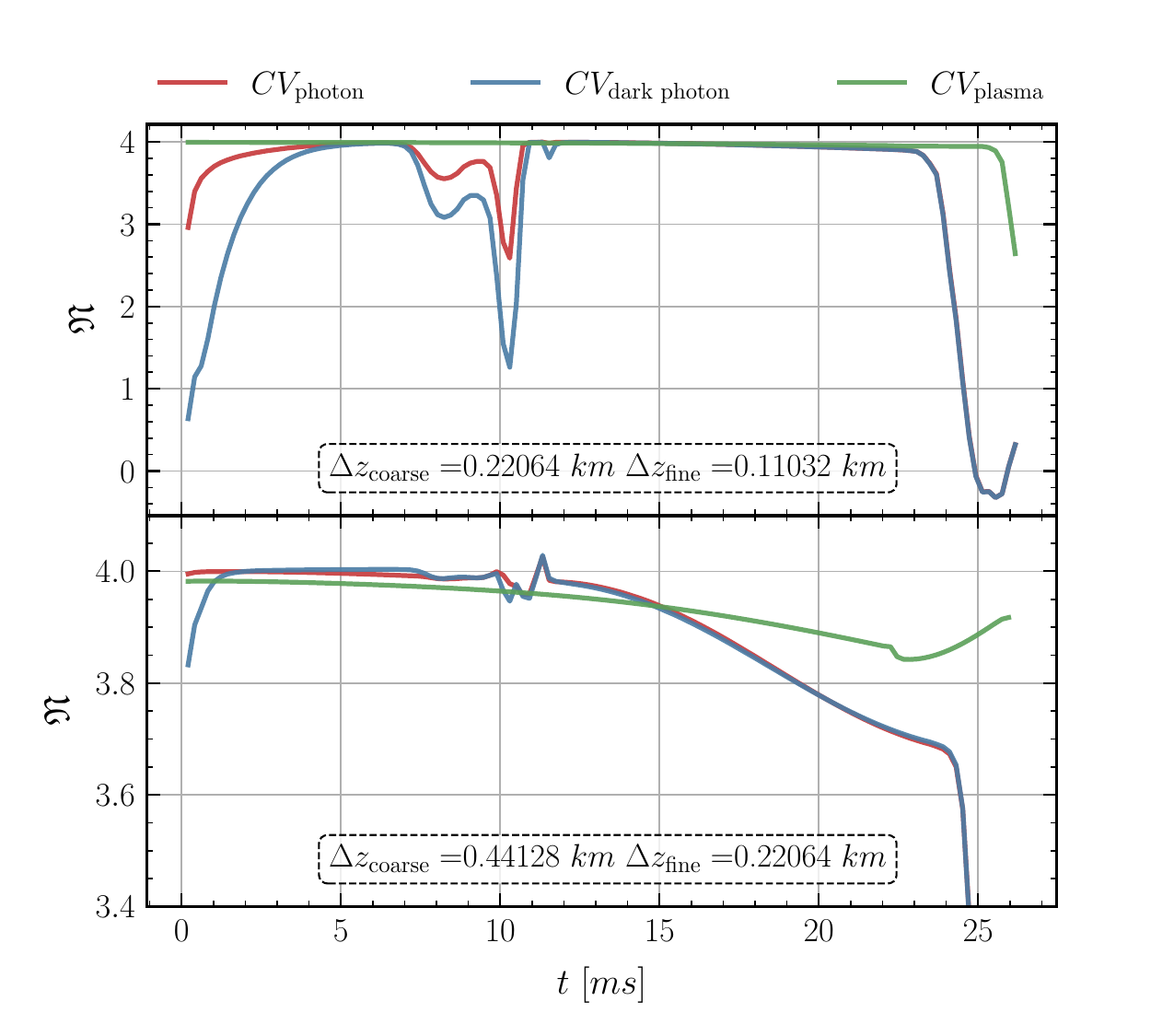}
    \caption{Convergence order for the simulations of the interaction of a wave packet of the electromagnetic field with a plasma barrier, in the regime where nonlinear effects appear. The computation is repeated for each constraint, and the results are shown with different colors. The upper and lower panels show the behavior of $\mathfrak N$ extracted from two pairs of simulations with higher and lower resolution, respectively. At $t \sim 1 \, \ms$ and $t \sim 10 \, \ms$ there are regions in which $CV_{\rm photon}$ and $CV_{\rm dark~photon}$ are more affected by numerical noise in the simulation with $\Delta z = 0.11032 \, \km$ than in the other cases, and $\mathfrak N$ is closer to 4 when computed from the simulations with lower resolution. For $t \gtrsim 10 \, \ms$  $\mathfrak N$ decreases slowly at first, and then convergence is completely lost before the crash of the simulations.
    \label{fig:NonLinearLZConvergenceOrder}
    }
\end{figure}

To study the global convergence we computed the volume-averaged L2 norm of the constraints as
\begin{equation}
    \lVert CV_{\Delta z} \rVert_2 = \sqrt{\frac{1}{z_{+\infty} - z_{-\infty}} \int_{z_{-\infty}}^{z_{+\infty}} dz \, CV_{\Delta z}^2},
    \label{eq:VolumeAveragedL2Norm}
\end{equation}
where $CV_{\Delta z}$ denotes a constraint violation among one of Eqs.~\eqref{eq:CVphoton}-\eqref{eq:CVplasma}, extracted from a simulation with grid step $\Delta z$. By computing this norm on both the simulations we could then obtain the convergence order $\mathfrak N$ as
\begin{equation}
    \mathfrak{N} = \log_{(\Delta z_{\rm coarse} / \Delta z_{\rm fine})} \frac{\lVert CV_{\Delta z_{\rm coarse}} \rVert_2}{\lVert CV_{\Delta z_{\rm fine}} \rVert_2}
\end{equation}
We evaluated this quantity on each of the constraints, estimating the integral in Eq.~\eqref{eq:VolumeAveragedL2Norm} with the trapezoidal rule, and we plotted the results in the upper panel of Fig.~\ref{fig:NonLinearLZConvergenceOrder}. At $t = 0 \, \ms$ all the constraint violations are identically zero, and in the initial stages they start to grow. While $CV_{\rm Plasma}$ immediately shows convergence, the other two constraints are initially dominated by round-off errors and numerical noise, and reach $\mathfrak N = 4$ more slowly. Approximately when the wave packet reaches the plasma barrier, the convergence order computed from $CV_{\rm photon}$ and $CV_{\rm dark~photon}$ drops, but then reaches the expected value again. After that it decreases mildly, and in the final stages before the simulation crashes convergence is lost again. By checking the behavior of the constraint violations at $t \sim 1 \, \ms$ and $t \sim 10 \, \ms$ we observed that there are regions in which $CV_{\rm photon}$ and $CV_{\rm dark~photon}$ are more affected by numerical noise in the simulation with higher resolution than in the lower resolution one. We therefore decided to perform an additional simulation with even lower resolution, $\Delta z = 0.44128 \, \km$, and we recomputed the convergence order $\mathfrak N$. The results are displayed in the lower panel of Fig.~\ref{fig:NonLinearLZConvergenceOrder}. Here we can see that in the early stages the convergence order is considerably closer to the expected value, while in the last stages the decrease in $\mathfrak N$ is more pronounced.

In conclusion, the test simulations in which there is a sufficient backreaction on plasma show that the implementation of the numerical evolution is correct, displaying good convergence properties. In the regime of interaction considered in this paper the expected scaling of the constraint violations with resolution is not satisfied. This is due to fact that the plasma density evolves on scales that mostly fall below machine precision and cannot be resolved by our code.

\bibliographystyle{bibi}
\bibliography{References}

\end{document}